\documentclass[twocolumn]{aastex631}

\newcommand\Hazelnut{\texttt{Hazelnut}\xspace}
\newcommand\lluvia{\texttt{lluvia}\xspace}

\usepackage{booktabs}
\usepackage{comment}
\usepackage{mhchem}
\usepackage{color}
\usepackage{upgreek} 
\usepackage{xspace} 

\begin{document}

\title{The HUSTLE Program: The UV to Near-IR Transmission Spectrum of the Hot Jupiter KELT-7b}

\author[0000-0001-5097-9251]{C. Gasc\'on}
\affiliation{Center for Astrophysics ${\rm \mid}$ Harvard {\rm \&} Smithsonian, 60 Garden St, Cambridge, MA 02138, USA}
\affiliation{Institut d'Estudis Espacials de Catalunya (IEEC), 08860 Castelldefels, Barcelona, Spain}

\author[0000-0003-3204-8183]{M. L\'opez-Morales} 
\affiliation{Space Telescope Science Institute, 3700 San Martin Drive, Baltimore, MD 21218, USA}

\author[0000-0003-4816-3469]{R. J. MacDonald}
\altaffiliation{NHFP Sagan Fellow}
\affiliation{Department of Astronomy, University of Michigan, 1085 S. University Ave., Ann Arbor, MI 48109, USA}
\affiliation{School of Physics and Astronomy, University of St Andrews, North Haugh, St Andrews, KY16 9SS, UK}

\author[0000-0003-3726-5419]{J. K. Barstow} 
\affiliation{School of Physical Sciences, The Open University, Walton Hall, Milton Keynes MK7 6AA, UK}

\author[0000-0002-4945-1860]{V. A. Boehm}
\affiliation{Department of Astronomy, Cornell University, 122 Sciences Drive, Ithaca, NY 14853, USA}

\author[0000-0003-4328-3867]{H. R. Wakeford}
\affiliation{School of Physics, University of Bristol, H.H. Wills Physics Laboratory, Tyndall Avenue, Bristol BS8 1TL, UK}

\author[0000-0003-4157-832X]{M. K. Alam}
\affiliation{Space Telescope Science Institute, 3700 San Martin Drive, Baltimore, MD 21218, USA}

\author[0000-0001-8703-7751]{L. Alderson} 
\affiliation{Department of Astronomy, Cornell University, 122 Sciences Drive, Ithaca, NY 14853, USA}

\author[0000-0003-1240-6844]{N. E. Batalha} 
\affiliation{NASA Ames Research Center, Moffett Field, CA 94035, USA}

\author[0000-0001-9665-5260]{C. E. Fairman}
\affiliation{School of Physics, University of Bristol, H.H. Wills Physics Laboratory, Tyndall Avenue, Bristol BS8 1TL, UK}

\author[0000-0001-5878-618X]{D. Grant} 
\affiliation{HH Wills Physics Laboratory, University of Bristol, Tyndall Avenue, Bristol, BS8 1TL, UK}

\author[0000-0002-8507-1304]{N. K. Lewis}
\affiliation{Department of Astronomy, Cornell University, 122 Sciences Drive, Ithaca, NY 14853, USA}

\author[0000-0002-5251-2943]{M. S. Marley}
\affiliation{Department of Planetary Sciences and Lunar and Planetary Laboratory, University of Arizona, Tuscon, AZ 85721, USA}

\author[0000-0002-6721-3284]{S. E. Moran}
\altaffiliation{NHFP Sagan Fellow}
\affiliation{Department of Planetary Sciences and Lunar and Planetary Laboratory, University of Arizona, Tuscon, AZ 85721, USA}
\affiliation{NASA Goddard Space Flight Center, 8800 Greenbelt Road, Greenbelt, MD 20771, USA}

\author[0000-0003-3290-6758]{K. Ohno} 
\affiliation{Department of Astronomy and Astrophysics, University of California, Santa Cruz, Santa Cruz, CA, USA}

\author[0000-0002-3645-5977]{G. Anglada-Escudé}
\affiliation{Institut de Ci\`encies de l'Espai (ICE, CSIC), Campus UAB, c/ de Can Magrans s/n, 08193 Bellaterra, Barcelona, Spain}
\affiliation{Institut d'Estudis Espacials de Catalunya (IEEC), 08860 Castelldefels, Barcelona, Spain}

\author[0000-0002-6689-0312]{I. Ribas}
\affiliation{Institut de Ci\`encies de l'Espai (ICE, CSIC), Campus UAB, c/ de Can Magrans s/n, 08193 Bellaterra, Barcelona, Spain}
\affiliation{Institut d'Estudis Espacials de Catalunya (IEEC), 08860 Castelldefels, Barcelona, Spain}

\begin{abstract}

The ultraviolet and optical wavelength ranges have proven to be a key addition to infrared observations of exoplanet atmospheres, as they offer unique insights into the properties of clouds and hazes and are sensitive to signatures of disequilibrium chemistry. Here we present the 0.2--0.8 $\micron$ transmission spectrum  of the $\rm T_{eq} = 2000\,K$ Jupiter KELT-7b, acquired with HST WFC3/UVIS G280 as part of the HUSTLE Treasury program. We combined this new spectrum with the previously published HST WFC3/IR G141 (1.1--1.7\,$\upmu$m) spectrum and ${\it Spitzer}$ photometric points at 3.6$\mu$m and 4.5$\mu$m, to reveal a generally featureless transmission spectrum between 0.2 and 1.7 $\micron$, with a slight downward slope towards bluer wavelengths, and a asymmetric water feature in the 1.1-1.7 $\mu$m band. Retrieval models conclude that the 0.2 -- 1.7$\mu$m  spectrum is primarily explained by a high \ce{H-} abundance ($\sim 10^{-5}$), significantly above the equilibrium chemistry prediction ($\sim 10^{-12}$), suggesting disequilibrium in KELT-7b's upper atmosphere. Our retrievals also suggest the presence of bright inhomogeneities in the stellar surface, and tentative evidence of ${\rm CO_2}$ at the ${\it Spitzer}$ wavelengths. We demonstrate that with the UV-optical coverage provided by WFC3 UVIS/G280, we are able to confirm the presence and constrain the abundance of \ce{H-}, and obtain evidence for  bright stellar inhomogeneities that would have been overlooked using infrared data alone. Observations redward of 1~$\micron$ with JWST should be able to further constrain the abundance of \ce{H-}, as well as confirm the presence of ${\rm CO_2}$ inferred by the two {\it Spitzer} datapoints.

\end{abstract}


\section{Introduction} \label{sec:intro}

Transmission spectroscopy of exoplanets over the past two decades --  mainly carried out with the \textit{Hubble Space Telescope} (HST) and large ground-based telescopes — has proven successful in studying the atmospheric composition of gas giant exoplanets at optical and near-infrared (NIR) wavelengths. These observations have yielded detections of atomic and molecular species such as Na {\sc i} and K {\sc i} in the optical \citep[e.g., ][]{sing2015,alam2018}, and \ce{H2O} in the NIR \citep[e.g., ][]{deming2013, wakeford2017, evans2018}.  JWST has recently started to improve and expand on these early results by obtaining transmission spectra of gas giant exoplanets between 0.6 and 15 $\mu$m. Highlights of recent JWST results on atmospheres of gas giant exoplanets include the detection CO$_2$, H$_2$O, CO, K {\sc i} and the photo-chemically produced SO$_2$ in the hot Jupiter WASP-39b, observed as part of the Early Release Science (ERS) program \citep{Ahrer2023, Alderson2023, Esparzaborges2023,Feinstein2023, Grant2023, rustamkulov2023, tsai2023, Powell2024}, and the detection of CH$_4$ in WASP-80b \citep{bell2023}.

Transmission spectroscopy observations have also revealed that most optical and NIR absorption features appear to be partially or completely muted, hinting at a prevalence of high altitude condensation clouds or photochemical hazes (aerosols) in the majority of exoplanet atmospheres \citep[e.g.,][]{sing2016, wakeford2020}. While still an open question, population studies have suggested a cloudy to clear transition at ${\rm T_{eq} \approx 2000\,K}$, believed to be produced by silicate and magnesium clouds condensing below this temperature, while above this temperature the chemical constituents of those clouds remain in the atmosphere in gas phase \citep{fu2017, tsiaras2018, lothringer2022}. However, the physical and chemical properties of these aerosols are poorly understood, and the intrinsic degeneracy between clouds and metallicity in transmission spectra models has challenged our ability to derive conclusions.

Observations blueward of 0.6 $\micron$ can help remove this major source of uncertainty by giving us access to the upper atmospheric layers of exoplanets, which can inform the presence of aerosols in a planet's atmosphere. By probing the scattering slope, these wavelengths can help constrain the particle size, composition, and vertical distribution of aerosols \citep[]{wakeford2015, ohno2020}. Furthermore, mid- to near-UV (M-NUV; 0.2 -- 0.4 $\micron$) wavelengths are particularly sensitive to signatures of chemical disequilibrium processes, such as the presence of elevated amounts of ionized hydrogen, H$^-$ \citep{lewis2020}, or silicate condensation precursors such as SiO \citep[e.g., ][]{evans2018, lothringer2022}. The atmospheres of a few exoplanets have been observed between 0.2 and 0.4 $\micron$ with HST \citep[e.g.,][]{wakeford2020, lothringer2022,Boehm2024}, but this number is still small, and the atmospheric properties of exoplanets at these wavelengths remain mostly unexplored. To address this observational gap, the Hubble Ultraviolet-optical Survey of Transiting Legacy Exoplanets (HUSTLE) Treasury Program (GO 17183, PI: Wakeford) is observing twelve gas giants with masses between $0.05 - 2.3 M_{J}$, radii between $0.6 - 2R_{J}$, and equilibrium temperatures between $900 - 2600$ K, to sample the potential cloudy-to-clear transition temperature regime, while also producing the largest UV-optical exoplanet survey to date. HUSTLE uses HST's Wide Field Camera 3 (WFC3) UVIS G280 grism, which is currently the most efficient instrument for obtaining UV-optical spectra of exoplanets \citep{wakeford2020,lewis2020,lothringer2022}.

Here we present the HUSTLE observations of KELT-7b. KELT-7b is a transiting hot Jupiter with a mass of $1.39 M_{J}$ and a radius of $1.6 R_J$ \citep{bieryla2015}, orbiting a $V = 8.6$, F-type star. With an equilibrium temperature of about 2040 K, KELT-7b is one of the hottest exoplanets within the HUSTLE sample and sits on the edge of the hot/ultra-hot Jupiter boundary. KELT-7b's atmosphere was studied by \cite{pluriel2020}, who analyzed transmission and emission spectra obtained with HST WFC3/IR G141 from $1.1$ to $1.7 \micron$. That transmission spectrum showed strong absorption features consistent with a cloud-free atmosphere containing both \ce{H2O} and \ce{H-}. Under thermochemical equilibrium, the hydrogen anion \ce{H-} is not expected to be significantly present in planets with $T < 2500$ K \citep{kitzmann2018}. However, disequilibrium processes such as photochemistry could increase the abundance of free electrons and neutral H at those temperatures by many orders of magnitude \citep{lavvas2014}, ultimately leading to the production of \ce{H-} via dissociative electron attachment at temperatures below 2500 K \citep{lewis2020}. The production of \ce{H-} could be particularly enhanced for hot Jupiters orbiting around F-type stars, as they receive high levels of extreme UV irradiation  \citep{lavvas2014, lewis2020}. This seems to be the case, for example, for WASP-17b \citep[1770K;][]{alderson2022}, WASP-79b \citep[1716K;][]{rathcke2021} and HAT-P-41b \citep[1950K;][]{wakeford2020, lewis2020}, all of which have shown some tentative signs of \ce{H-} in their atmospheres. KELT-7b is another such planet, with an effective temperature below 2500 K around an F-type star, and thus a suitable target to test whether or not the atmospheres of this type of planets appear H$^-$ enriched.

The paper is organized as follows. In Section~\ref{sec:obs}, we describe our KELT-7b observations and the data reduction process. Section~\ref{sec:lcfit} describes the fitting of the light curves and presents the corresponding transmission spectrum. In Section~\ref{sec:retrievals} we describe the atmospheric retrievals and discuss the results and implications. In Section~\ref{sec:comp} we compare our observations of KELT-7b to the spectra of similar hot Jupiters with published UV-optical observations. We present our conclusions in Section~\ref{sec:conclusions}. 

\section{Observations and Data Reduction} \label{sec:obs}

\subsection{Observations}

We observed one transit of KELT-7b with the HST WFC3/UVIS G280 grism between 29 December 2022 14:19:42 and 21:42:13 UT.  The visit consisted of 127 exposures of $16$ seconds each distributed over five HST orbits, giving approximately 25 exposures per orbit. We used the 800$\times$2100 pixel subarray centered on chip 2, which reduced the readout times while still capturing both the +1 and -1 spectral orders around the zeroth order trace (see Figure~\ref{fig:wlc}). To accurately measure the position of the target on the detector subarray and aid in the wavelength calibration, we collected a 1.0 s direct image with the F300X filter at the start of the observation. 

\subsection{Data Reduction}

To guarantee the robustness of our results, we analyzed the data using two independent pipelines, called  \texttt{lluvia} and \texttt{hazelnut} \citep{Boehm2024}. Both reductions started the analysis on the \texttt{flt} files generated by the STScI \texttt{calwf3} pipeline (v.3.6.2), which applies standard calibrations to the raw data, i.e. dark subtraction, flat fielding and bias corrections. Both pipelines also re-embed the subarray images into the full frame using the \texttt{spt} files and the \texttt{embedsub} routine available in the \texttt{wfc3tools} package\footnote{\url{https://github.com/spacetelescope/wfc3tools}}. After that, the reductions proceeded independently as described below. 

\subsubsection{\lluvia reduction}

\begin{figure*}
    \centering
    \includegraphics[width = \textwidth]{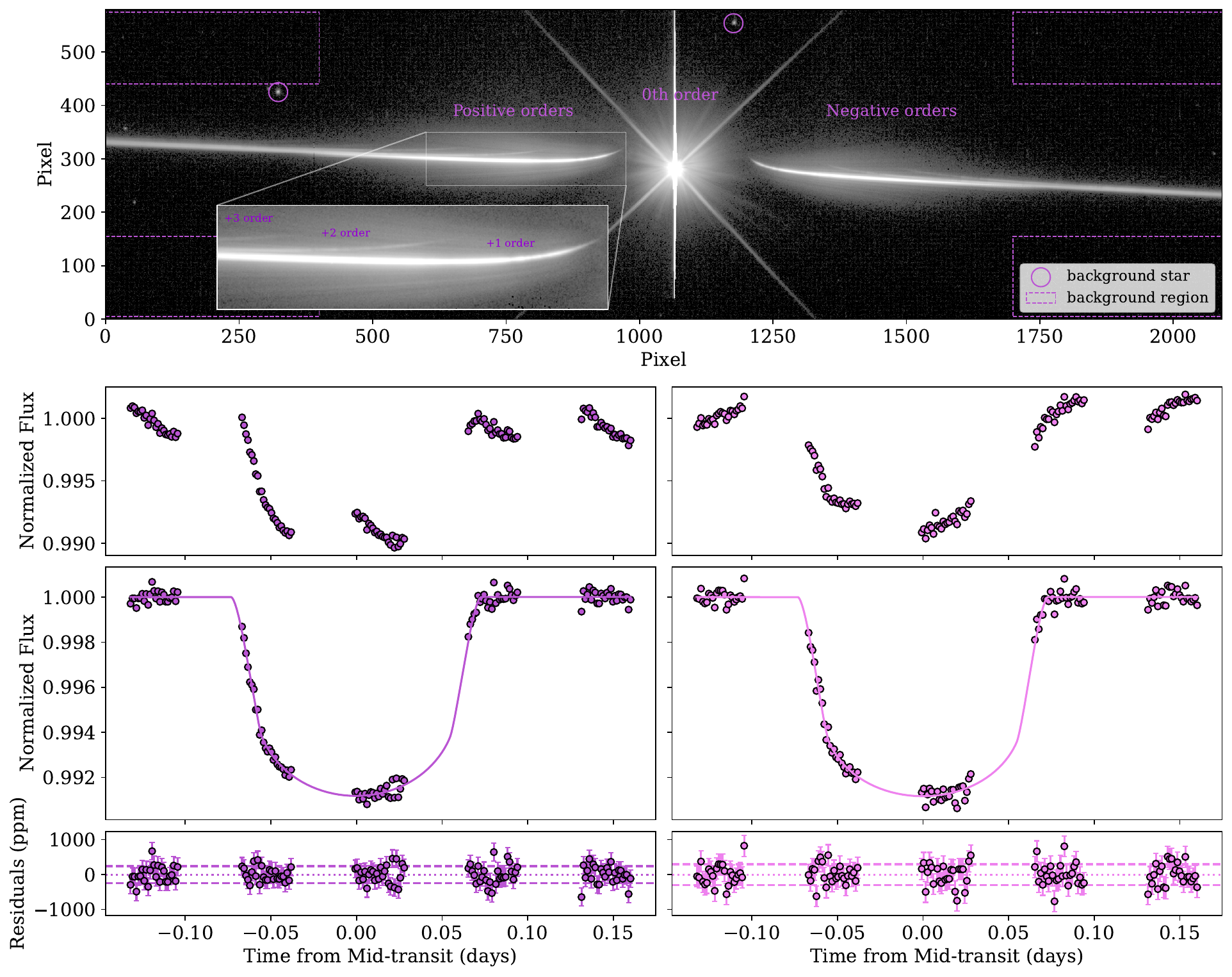}
    \caption{Top: HST WFC3/UVIS G280 exposure showing the order zero image, and the spectra corresponding to positive and negative orders. Two background stars are highlighted by purple circles. The dashed rectangles in the image corners indicate the region used to estimate background on each image. Middle: Raw +1 (left) and -1 (right) order white-light curve obtained with \lluvia. Bottom: Systematics corrected +1 (left) and -1 (right) order white-light curve. The narrow panel below shows the transit fit residuals, with the dashed lines indicating the corresponding standard deviation of the residuals.}
    \label{fig:wlc}
\end{figure*}

We estimated the background flux of each exposure using 150$\times$400 pixel regions located in the four corners of the corresponding image, as illustrated by the purple dashed rectangles in the top panel of Figure~\ref{fig:wlc}. We used the central value of the combined histogram representing the distribution of counts in those four regions as the image background, which we then subtracted from each exposure. In the case of KELT-7b, those central values, determined by fitting a Gaussian to each image's histogram, were negative (between -8 and -5 $e^-$), which we attribute to an inaccurate bias-correction from \texttt{calwf3}. We also performed an initial cosmic ray correction on each image by iteratively identifying pixels that deviated by more than 5$\sigma$ in time within each HST orbit, and replacing their flux values by the median pixel value of the corresponding orbit. 

We computed the expected trace location of both +1 and -1 orders, together with the corresponding wavelength solutions, using the UVIS G280 calibration presented in \cite{pirzkal2020}, and implemented in the python package \texttt{grismconf}\footnote{\url{https://github.com/npirzkal/GRISMCONF}}. In their work, \cite{pirzkal2020} describe the trace location as a 6th order polynomial, with coefficients determined by a second order 2D field polynomial. This 2D polynomial is solely a function of the target’s $X$ and $Y$ coordinates in the detector, which we determined by calculating the centroid of KELT-7 in the F300X direct image taken at the start of the observations. To account for any drifts or displacements of the spectra on the detector over time, we refined the position of the spectral trace by fitting a Gaussian profile to each column of the detector (i.e., in the cross-dispersion direction), using the calibration trace as an initial guess. For each exposure we found excellent agreement between the two computed traces, reflecting both the accuracy of the calibration from \cite{pirzkal2020}, and HST's stability throughout the observation. Tracking the center of the Gaussian profile in time already gives us an estimation of the vertical displacement of the image along the observation. We however compute this displacement more accurately via cross-correlation of the cross-dispersion profile, as described later in this section.

\begin{figure*}
    \centering
    \includegraphics[width = \textwidth]{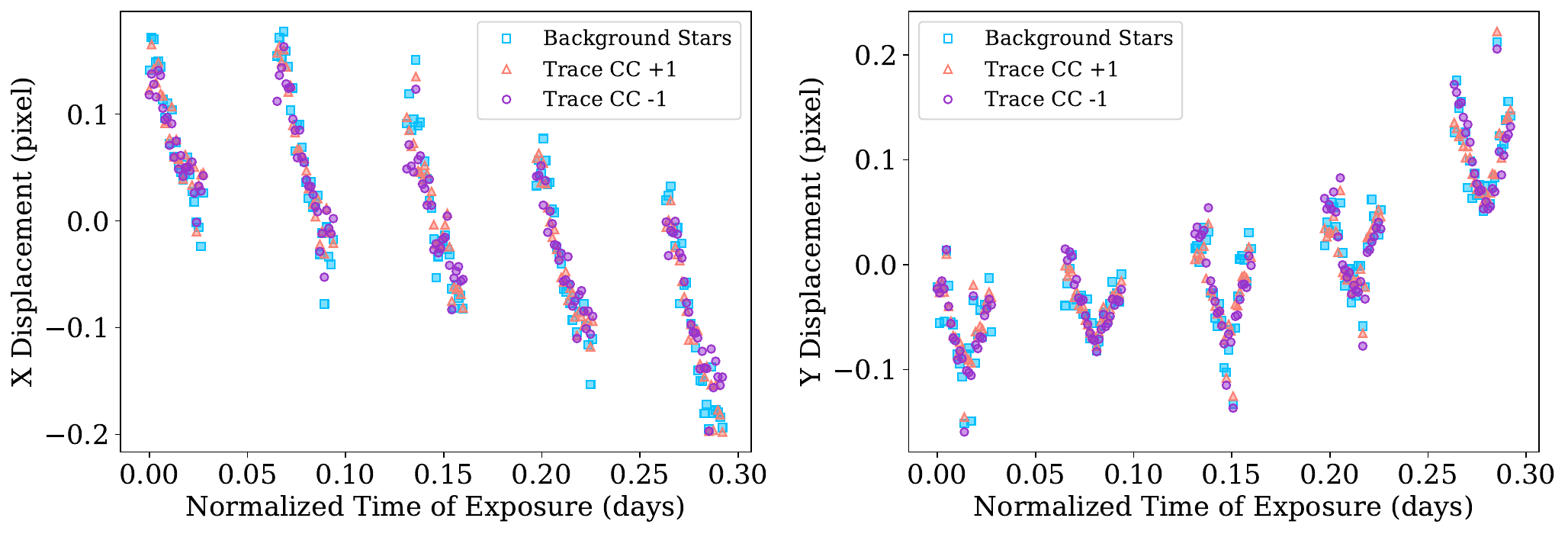}
    \caption{Displacements in the X (left) and Y (right) detector directions, measured by cross-correlation of the spectral trace (red triangles and purple circles), and by tracking the centroid position of several background stars (blue squares).}
    \label{fig:disps}
\end{figure*}

We extracted the stellar spectra using the optimal extraction routine described in \cite{Horne1986}. Similarly to \cite{Marsch1989}, we built the spatial profile by fitting low-order polynomials along the trace direction, hence accounting for the characteristic curvature in the trace of UVIS G280 observations (see Figure~\ref{fig:wlc}). During this fitting process, we performed a second outlier removal in which we identified pixels that deviated from this spatial polynomial fit by more than $7\sigma$, and replaced them by the corresponding value of the polynomial fit. We subsequently extracted the stellar spectra using a $\pm15$ pixel width aperture centered on the fitted trace, which we found to be optimal for extracting both +1 and -1 orders as it minimized the white light curve residuals. Finally, we measured the shift of the trace in the dispersion direction $\delta_\lambda$  by cross-correlating the extracted spectra with a template median spectrum. We subsequently used $\delta_\lambda$ to shift and align all the spectra. Similarly, we measured the shifts in the cross-dispersion direction ($Y_{\text{psf}}$) by cross-correlating the trace profiles with a template median profile. Figure~\ref{fig:disps} shows the resulting X and Y displacements obtained by cross-correlation of the +1 and -1 order profiles (red and purple markers, respectively). We also tracked the centroid position of two unsaturated background stars in the image (indicated by purple circles in Figure~\ref{fig:wlc}), which yielded very similar, but less precise displacements than those computed with the trace cross-correlation (Figure~\ref{fig:disps}, blue squared markers). The values of $\delta_\lambda$ and $Y_{\text{psf}}$ were used as detrending variables during the light curve fitting process, as described in Section~\ref{sec:lcfit}.

We generated a white light curve for each order by integrating the flux in the extracted spectra between 200 and 750 nm. The +1 and -1 raw light curves, after removal of outliers in flux, are shown on the left and right middle panels of Figure~\ref{fig:wlc}, respectively. We also generated 57 spectroscopic light curves between 200 and 750 nm, using 10 nm bins. The +1 order raw spectroscopic light curves are shown in the left-side column of Figures \ref{fig:lcs1} and \ref{fig:lcs2}, in Appendix~\ref{sec:anexlcs}. Similar spectroscopic light curves were obtained for the -1 order.

\subsubsection{\Hazelnut reduction} \label{subsubsec:hazelnutred}

In contrast to our \lluvia analysis, our \Hazelnut reduction began with the removal of the first frame of each orbit from further analysis. This decision was motivated by previous reports that the first frame in each orbit can be more strongly affected by unique systematics and excess earthshine \citep{wakeford2020}; however, we find in the case of KELT-7b that first-frame removal makes no significant difference in the results. We next removed cosmic rays by iterating twice over the time series of each pixel, replacing 4.5$\sigma$ outliers with the median of six adjacent values in time. We lastly removed the sky background signal by computing the mode of each full frame and subtracting this value from all pixels in the frame. We find that the histograms of pixel values in our frames reveal a bimodal distribution of values; many pixels were valued between -8 to -5 e-, but a similar number of pixels had higher values of around 2 e-. In some frames, slightly more pixels had a value of 2 e- so that this value, rather than the negative mode, was taken as the background signal, resulting in an artificial ``flickering'' after background subtraction. The higher-valued pixels were found to cluster around the edges of the 0th order, while the lower-valued pixels were more evenly dispersed across the frame. We concluded from these spatial patterns that the higher mode was the result of diffusion of 0th order signal across the frame, while the lower mode was the true background signal. We thus treated the lower-valued mode, which ranged between -8 to -5 $e^-$ as in \lluvia, as our background signal and removed this value from each frame, eliminating the flickering systematic signal from our reduction.

As with \lluvia, we fit the locations of the +1 and -1 traces using \texttt{grismconf} \citep{pirzkal2020}. Our extraction used a standard, unweighted aperture of halfwidth 10 pixels, which was selected as it minimized the scatter in the out-of-transit residuals. Pointing instabilities during observation result in subpixel shifts of the traces across the detector over time, which can lead to discrepancies between the spectra and the wavelength solution produced by \cite{pirzkal2020}. To correct for these shifts, we cross-correlated all spectra with the spectrum extracted from frame 2 to measure the drift in the dispersion direction, and used the measured drift to align the spectra with the frame 2 spectrum. An additional round of cosmic ray rejection was performed at this stage, in which the timeseries of each point in the 1D extracted spectra were iterated to remove 3.5$\sigma$ outliers, which were subsequently replaced by the median value in time. Our pipeline likewise generated a white light curve for each order, summing the flux from 200~nm to 800~nm. We generated 60 spectroscopic light curves per order in 10~nm bins spanning 200~nm to 800~nm, for a total of 2 white light curves and 120 spectroscopic light curves. The +1 and -1 raw white light curves are shown on the left and right top panels of Figure~\ref{fig:wlc_abby}, respectively.

\section{Light curve fitting}
\label{sec:lcfit}

\begin{figure*}
    \centering
    \includegraphics[width = \textwidth]{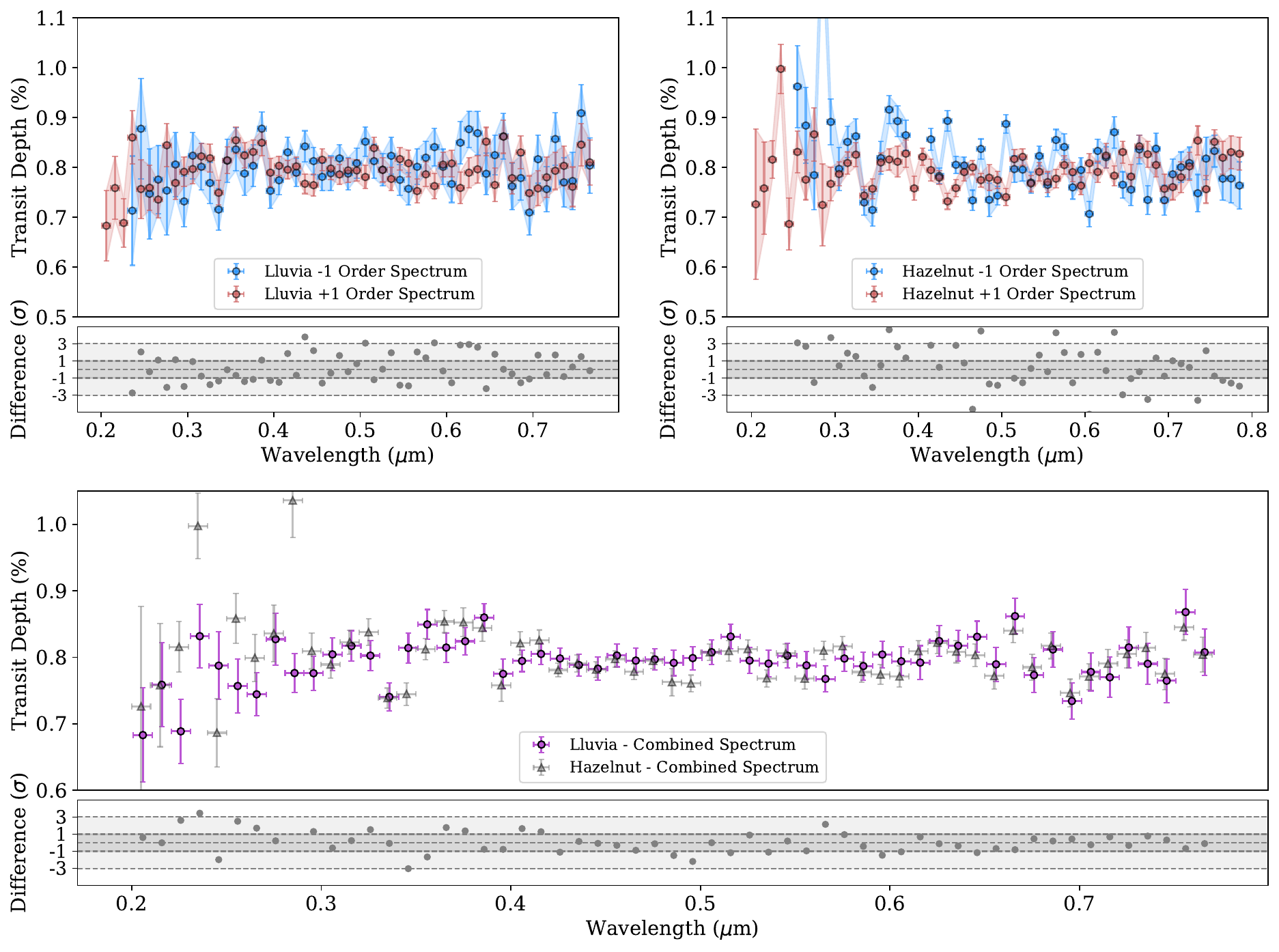}
    \caption{Results from the KELT-7b light curve fitting. Top: +1 (blue) and -1 (red) order transmission spectra obtained with the \lluvia pipeline (left), and the \Hazelnut pipeline (right). Bottom: Combined transmission spectrum obtained with \lluvia (purple), and the \Hazelnut (gray). The narrow panels below each plot show the differences between the corresponding transmission spectra in units of $\sigma$, with the dashed lines and the shaded regions indicating the corresponding $\pm1\sigma$ and $\pm3\sigma$ limits.}
    \label{fig:transm_spec}
\end{figure*}

Each pipeline fitted the extracted white and spectroscopic light curves for orders +1 and -1 independently, following similar procedures to the ones outlined in \cite{sing2019} and \cite{wakeford2020}.

\subsection{\lluvia}

For each light curve extracted with \lluvia, we modeled the flux measurements over time $f(t)$ using a function of the form
\begin{equation}
    f(t) = F_{0} \times T(t, \theta) \times S(x) ,
\end{equation}
\noindent where $F_{0}$ is the baseline flux, $T(t, \theta)$ is the analytic transit model as implemented in the \texttt{batman} package \citep{kreidberg2015}, $\theta$ are the orbital parameters defining the transit model, and $S(x)$ is an instrument systematics model that uses optical state vectors extracted from the observations and jitter vectors from the HST Pointing Control System \citep{sing2019}. The full functional form of our systematic model is
\begin{eqnarray}\label{eq:sys}
    S(x) = p_{1}\phi_{t} + \sum_{i = 1}^{4}p_{i+1}\phi_{\text{HST}}^{i} + p_{6}\delta_{\lambda} + p_{7}\sigma_{\text{prof}}\nonumber\\
    + p_{8}Y_{\text{psf}} + p_{9}\text{R.A.} + p_{10}\text{decl.} \nonumber\\
    + p_{11}V2_{\text{roll}} + p_{12}V3_\text{roll} + 1,  
\end{eqnarray}
\noindent where $\phi_t$ is a linear baseline trend, $\phi_{\text{HST}}$ is \textit{HST}'s orbital phase, $\delta_{\lambda}$ is the wavelength shift of each spectrum, $\sigma_{\text{prof}}$ is the mean profile width, $Y_{\text{psf}}$ is the image displacement along the cross-dispersion direction,  R.A.  and  decl. are the right ascension and declination respectively, $V2_\text{roll}$ and $V3_\text{roll}$ are the roll of the telescope along the $V2$ and $V3$ axes, and $p_{i}$ are the fitting coefficients for each detrending variable. For all light curves we performed an initial fit using a least squares minimization algorithm, which we used to reject 3.5$\sigma$ outliers and rescale the data point error bars so that the reduced chi-squared equaled unity. We then performed second fits using Markov Chain Monte Carlo (MCMC) with \texttt{emcee} \citep{foreman2013}. For each MCMC fit, we ran 32 walkers with 5000 steps each, 500 of which were discarded as burn-in. 

We first modeled the white light curves, fitting only for the mid-transit time, $T_0$, and the planet-to-star radius ratio, $R_p/R_s$. Because of the sparse phase coverage of HST observations caused by Earth's occultation, we fixed the semi-major axis, $a/R_s$, the orbital inclination, $i$, the orbital period, $P$, the eccentricity $e$, and the argument of the periastron, $\omega$, to the values used in the analysis of \cite{pluriel2020} -- see their Table 1. In addition to the parameters above, we also adopted a four-parameter non-linear limb darkening law \citep{sing2010, claret2000}, with coefficients computed with ExoTiC-LD \citep{grant_2022}, using the 3D stellar models from \cite{magic2015} and the stellar parameters taken from \cite{bieryla2015}. We fitted the transit model simultaneously with the systematic model in Equation~\ref{eq:sys}, testing all the possible combinations of the 12 detrending variables considered, and using the second order Akaike Information Criterion ($\text{AIC}_c$) to determine the best fitting subset of detrending variables. The resultant detrended white light curves for each spectral order are shown in the bottom panels of Figure~\ref{fig:wlc}, with a residuals standard deviation of 242 ppm for the +1 order and 300 ppm for the -1 order, respectively. From these fits, we measure a combined $R_p/R_s$ of 0.08932 $\pm$ 0.00026, and $T_0$ = 2459943.24192 $\pm$ 0.00017 BJD TDB, which is in good agreement with previously published ephemerides. The corresponding white light curve transit depth is 0.7978 $\pm$ 0.0046 \%. This is in good agreement with the transit depth measured with TESS (0.7961 $\pm$ 0.0018 \%, \cite{pluriel2020}), which was obtained in a similar bandpass as WFC3/UVIS G280. These values however slightly differ from the transit depths measured with the Spitzer 3.6$\micron$ channel (0.7925 $\pm$ 0.0062 \%) and the 4.5 $\micron$ channel (0.8092 $\pm$ 0.0036 \%), demonstrating the variation in transit depth as a function of the wavelength bandpass.

For the spectroscopic light curves, we performed a similar analysis, fixing $T_0$ to the value derived above and only fitting for $R_p/R_s$. For each order, we used the best-fitting form of Equation~\ref{eq:sys} derived from the corresponding white light curve analysis.  We achieved near to photon-noise precision, with only 7 out of the 114 light curves analyzed (57 for each order) needing an error bar rescaling of more than 20\%.  We also looked for any uncorrected systematics in the light curves by applying the binned residuals technique \citep{Pont2006}, finding no clear evidence of red noise in the residuals.  We extracted two independent transmission spectra for orders $+1$ and $-1$, which we subsequently combined with a weighted mean to produce the final transmission spectrum. The results obtained with the \lluvia pipeline are presented in Figure~\ref{fig:transm_spec}, with the upper-left panel showing the separate +1 and -1 order spectra, and the lower panel showing the combined spectrum in purple. The values of $(R_p/R_s)^2$ for each bin, together with the corresponding uncertainties, are listed in Table~\ref{tab:transmspec}. Furthermore, Figures \ref{fig:lcs1} and \ref{fig:lcs2} (Appendix~\ref{sec:anexlcs}) show the full set of raw and systematics corrected light curves for the +1 order. We obtained similar results for the -1 order.

\subsection{\Hazelnut}

We modeled the white light curves and the spectroscopic light curves produced by \Hazelnut in two steps. We first modeled the systematics in the light curves using the ExoTiC-ISM package \citep{Exoticism}, the procedure behind which is detailed in \cite{Wakeford2016}. ExoTiC-ISM fits a weighted combination of 51 systematics models to remove known trends in \textit{HST}, including breathing, visit-long slopes, and ramp effects. All white light and spectroscopic light curves were detrended in this way. We then fit the detrended light curves with an analytic transit model implemented through \texttt{batman} \citep{kreidberg2015}, with a four-parameter nonlinear stellar limb darkening law using coefficients generated by ExoTiC-LD \citep{grant_2022} from the 3D stellar model grid of \cite{magic2015}, and with stellar parameters sourced from \cite{bieryla2015}. We first perform linear least-squares fits to estimate the scatter and remove 3$\sigma$ outliers from each light curve, and then perform subsequent \texttt{emcee} MCMC fits to extract our final parameters with uncertainty. We initialized MCMC with the parameters from \cite{bieryla2015} and ran 32 chains with 5000 steps each, discarding the first 1000 steps (20\% of each chain) as burn-in.

We began by fitting the +1 and -1 white light curves to estimate the mid-transit time, $T_0$, and planet-to-star ratio, $R_p/R_s$, with other system parameters (i.e. $a/R_s$, $i$, $e$, etc.) fixed to the values used in \cite{pluriel2020}. Our fits achieved a residuals standard deviation of 374 ppm and 452 ppm in the +1 and -1 order, respectively (Figure~\ref{fig:wlc_abby}), with $R_p/R_s=0.08909\pm 0.00027$ and $T_0=2459943.24114\pm 0.00004$ BJD TDB, red which agrees within 4.5$\sigma$ with the $T_0$ obtained with \lluvia. The difference in the precision achieved between both pipelines may arise from the different detrending and fitting routines employed. For instance, the amount of vectors used during jitter decorrelation with \lluvia can increase the uncertainties compared to the system marginalization employed with \Hazelnut. Our spectroscopic light curve analysis followed a similar procedure, but with $T_0$ locked to the best-fit value obtained by our white light curve fits. The transmission spectrum obtained by this second analysis is presented in Figure~\ref{fig:transm_spec}, with the upper-right panel showing the separate +1 and -1 order spectra, and the lower panel showing the combined spectrum in gray. The resulting fits to the +1 order spectroscopic light curves and their residuals are shown in Figure~\ref{fig:lcs3}, in Appendix~\ref{sec:anexlcs}. We obtained similar fits for the spectroscopic light curves of order -1.

\subsection{Light curve fitting results}

The final transmission spectra obtained with both \lluvia and \Hazelnut are plotted together in the lower panel in Figure~\ref{fig:transm_spec}, with the bottom panel showing the difference between the two extractions. The results show a relatively flat spectrum for wavelengths above $0.4 \micron$, with a downward slope below $0.3 \micron$, and with no evident absorption features from alkali metals such Na or K, or from gaseous refractory species such as Fe, Mg or SiO. Both extracted spectra show good agreement especially towards the redder end of the spectrum, with 60\% of the points within 1$\sigma$ and 95\% of the points within a 3$\sigma$ uncertainty range. As evidenced in Figures \ref{fig:wlc} and \ref{fig:wlc_abby}, the fits conducted with \lluvia yield lower deviations in the white light curve residuals. These differences could be associated with several stages along the data reduction process, such as the background removal strategy, or the aperture size and extraction routine (optimal vs box) used by each pipeline. The light curve detrending strategy might have also been more effective in the case of the \lluvia reduction due to the increased amount and variety of jitter vectors used. Furthermore, the +1 and -1 order transmission spectra extracted with \lluvia show a better agreement than \Hazelnut, with 94\% of the points deviating by less than $3\sigma$ between both orders. We therefore decide to proceed with the \lluvia results for the atmospheric retrievals presented in the upcoming sections.

\section{Atmospheric Retrievals} \label{sec:retrievals}

We fitted the transmission spectrum of KELT-7b using two independent retrieval codes, as described below. In both cases, we performed the retrievals on the WFC3/UVIS G280 data from this work obtained with the \lluvia pipeline, together with the WFC3/IR G141 spectrum from \cite{pluriel2020} and the Spitzer data from \cite{baxter2021}. For each retrieval code, we performed an additional analysis including also the TESS data point from \cite{pluriel2020}.

\begin{figure*}
    \centering
    \includegraphics[width = \textwidth]{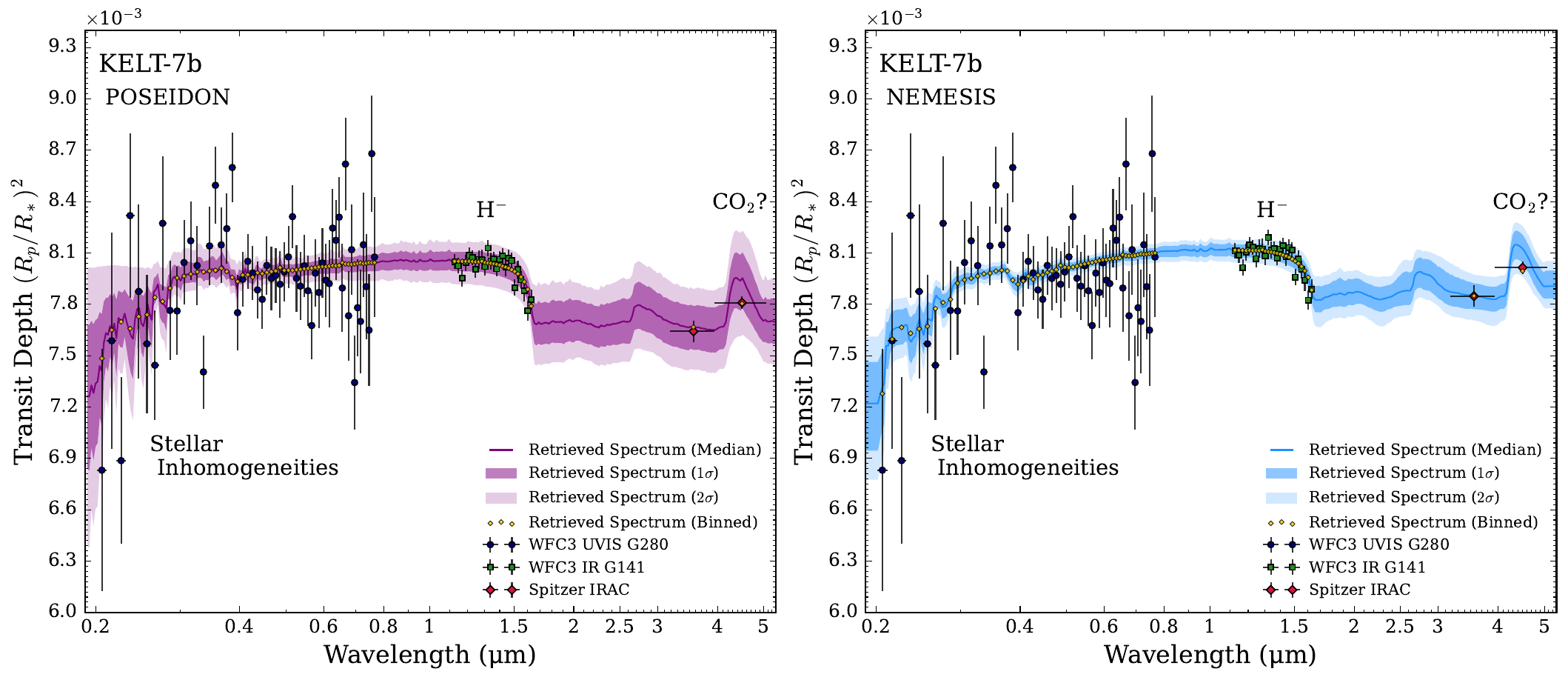}
    \caption{Atmospheric retrievals of KELT-7b's transmission spectrum conducted with \textsc{POSEIDON} (left panel, purple) and NEMESIS (right panel, blue). Each panel shows the median retrieved spectrum (solid lines), together with the 1$\sigma$ (dark shading) and 2$\sigma$ (light shading) regions.}
    \label{fig:ret_spectra}
\end{figure*}

\begin{figure*}
    \centering
    \includegraphics[width = \textwidth]{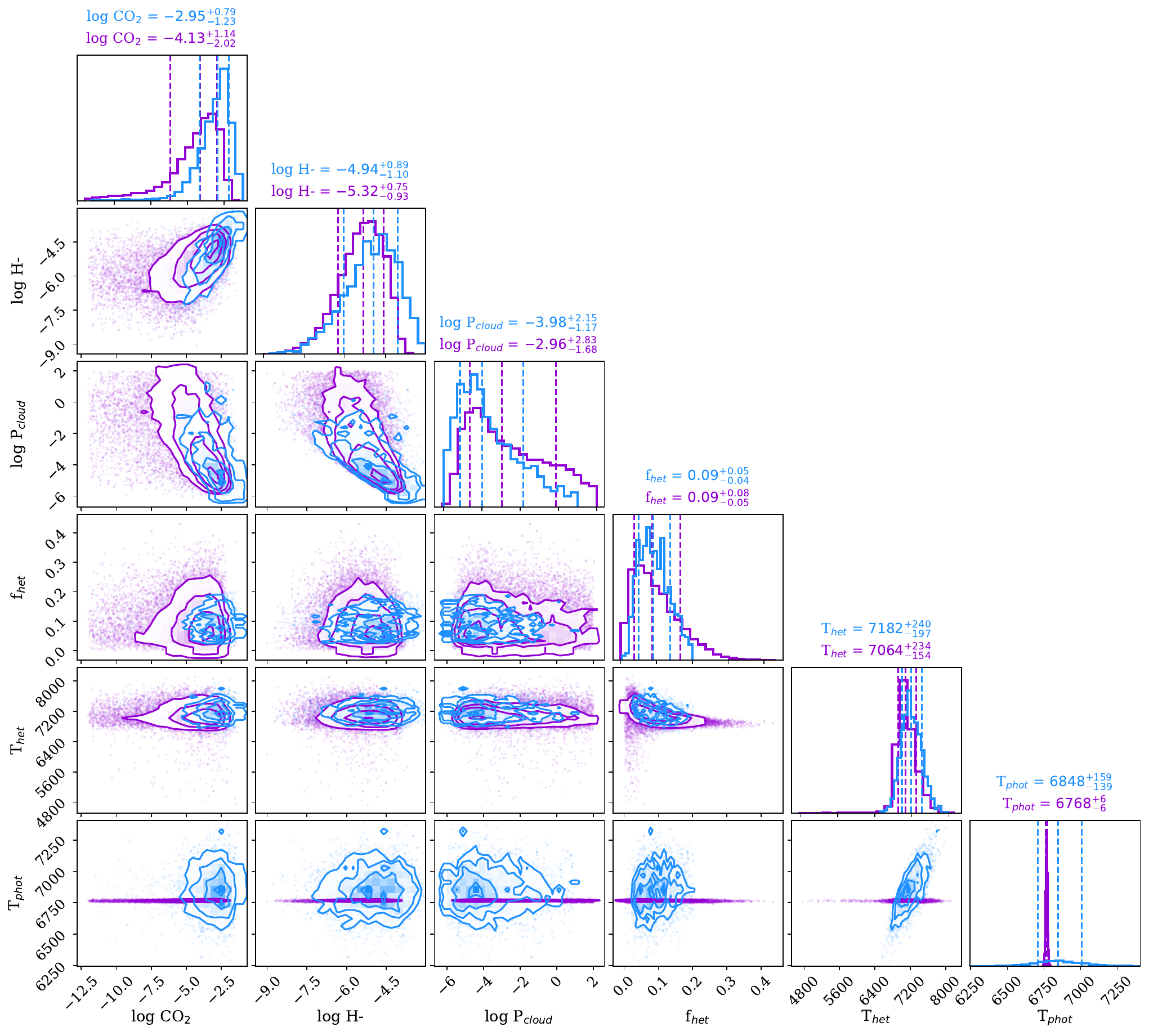}
    \caption{KELT-7b atmospheric retrieval posterior distributions of a subset of the parameters retrieved with \textsc{POSEIDON} (purple) and \textsc{NEMESISPY} (blue). We show constraints on the CO$_2$ and H$^-$ abundance, the cloud top pressure, the temperature and covering fraction of the stellar heterogeneities, and the temperature of the photosphere. While \textsc{POSEIDON} uses a much tighter prior on $T_{\rm{phot}}$ than \textsc{NEMESISPY}, resulting in a sharp distribution, the results are nevertheless consistent.}  
    \label{fig:corners}
\end{figure*}

\subsection{\textsc{NEMESISPY} retrieval setup} \label{sec:NEMESIS}

We performed a series of retrievals with the \textsc{NEMESISPY} radiative transfer and retrieval tool \citep{irwin2008,yang2024} coupled with the PyMultiNest nested sampling solver \citep{feroz08,feroz09,Feroz2019,Buchner2014,krissansen18} using 1,000 live points. We include absorption by H$_2$O \citep{Polyansky2018H2O}, CO$_2$ \citep{yurchenko2020}, CO \citep{Li2015}, TiO \citep{mckemmish19}, VO \citep{mckemmish19}, AlO \citep{patrascu2015}, SiO \citep{barton13}, FeH \citep{wende10}, SH \citep{gorman19} and H$^-$ \citep{john1988} (as they are expected to be the most abundant at the temperature of KELT-7b), and retrieve the volume mixing ratio for each of these gases. We use correlated-k tables for all except H$^-$, which is a continuum opacity feature, and these tables are generated according to \cite{chubb21}. H$_2$ and He continuum opacity is also included \citep{borysow89,borysowfm89,borysow90,borysow97,Borysow2001,borysow02}, and we assume an He/H$_2$ ratio of 0.17. In addition to the gas volume mixing ratios, we also retrieve the temperature-pressure profile using the \cite{Madhusudhan2009} parameterization. We also include a parametric cloud following \cite{MacDonald2017a} and \cite{Barstow2020a}; we retrieve a grey cloud top pressure, and a scattering index and opacity scaling for a haze lying above the grey cloud. We retrieve the radius of the planet at a pressure of 10 bar, which is the bottom of our model atmosphere. We also include a parameterisation for the effect of inhomogeneities in the stellar surface, such as those produced by unocculted starspots or faculae. This is parameterised as in \cite{lewis2020} and \cite{rathcke2021}, with the free parameters being the temperature of the unspotted photosphere, the difference in temperature between the stellar heterogeneities and the photosphere (negative for spots cooler than the photosphere, positive for regions hotter than the photosphere) and the heterogeneity covering fraction of the visible stellar disc. We calculate the photospheric, spot and facula spectra using stellar models from the \cite{castelli2003} catalogue and the pysynphot package by the \cite{pysynphot}. We assume a stellar log(g) of 4.15 and a metallicity of 0.139, following \cite{bieryla2015}, with the photospheric temperature allowed to vary in the retrieval. Given that the retrieval only fits for potential colder or hotter regions in the stellar surface, and not for actual physical processes like those leading to the onset of stellar faculae, we will use the term ``stellar inhomogeneities" throughout the paper. This is further discussed in Section~\ref{sec:faculae}. Finally, we include offsets between the WFC3/UVIS G280 and WFC3/IR G141 data and between the WFC3/UVIS G280 and Spitzer data as free parameters in the retrieval since these measurements were obtained several years apart, and we therefore cannot assume a constant baseline. Being only two individual data points, we do not add an offset between the two Spitzer channels since the inclusion of an offset would make the Spitzer data uninformative. Furthermore, if the systematics are appropriately dealt with, it is common practice to trust the spectral shape provided by the two Spitzer channels \citep[e.g., ][]{wakeford2017_2,alderson2022}. We list the priors for each free parameter in Table~\ref{tab:retrievals}.

\subsection{\textsc{POSEIDON} retrieval setup}
\label{sec:poseidon}

We also retrieved the KELT-7b UV to IR transmission spectrum described above with the open-source\footnote{\url{https://github.com/MartianColonist/POSEIDON}} \textsc{POSEIDON} retrieval package \citep{MacDonald2017a,MacDonald2023}. \textsc{POSEIDON} uses the nested sampling package PyMultiNest \citep{feroz08,feroz09,Feroz2019,Buchner2014} to explore the parameter space with 2,000 live points. We calculate model transmission spectra for each combination of atmospheric properties at $R =$ 10,000 from 0.19--5.4\,µm, before convolving and binning the model spectra to the resolution of the HST and \textit{Spitzer} data.

Our \textsc{POSEIDON} retrieval model considers several potentially important opacity sources, non-isothermal pressure-temperature profiles, parameterized aerosols, stellar contamination, and instrumental offsets. We assume H$_2$-He dominated atmospheres (using a fixed ratio of He/H$_2$ = 0.17) and consider a range of trace chemical species with strong cross sections over the wavelength range of our HST and \emph{Spitzer} observations. Specifically, we consider 10 chemical species parameterized by their $\log_{10}$ volume mixing ratios (line lists in parentheses): H$_2$O \citep{Polyansky2018H2O}, CO$_2$ \citep{yurchenko2020}, CO \citep{Li2015}, TiO \citep{mckemmish19}, VO \citep{mckemmish19}, AlO \citep{patrascu2015}, SiO \citep{barton13}, FeH \citep{wende10}, SH \citep{Yurchenko2018} and H$^-$ \citep{john1988}. Each model atmosphere is subdivided into 100 layers spaced uniformly in log-pressure from 10$^{-7}$ -- 10$^{2}$\,bar. We treat aerosols as a power-law haze and opaque cloud deck via 3 free parameters \citep{MacDonald2017a}, assuming, for simplicity, uniform aerosol coverage around the terminator. We marginalize over possible stellar contamination from unocculted starspots or faculae by interpolating PHOENIX models \citep{Husser2013} using the PyMSG package \citep{Townsend2023} and the 3-parameter stellar contamination prescription from \citet{rathcke2021}. We fit for non-isothermal pressure-temperature profiles using an adaptation of the 6-parameter profile from \citet{Madhusudhan2009} with the reference temperature set at 10\,mbar. We solve hydrostatic equilibrium assuming the ideal gas law and using a boundary condition determined by fitting for the reference planetary radius at 10\,bar. Finally, we add two free offsets, one between the WFC3/IR G141 data and the WFC3/UVIS G280 data and one between the \textit{Spitzer} IRAC data and WFC3/UVIS G280, for a total of 25 free parameters. We ran retrievals both with and without the TESS data point, without an additional offset for the TESS data. The priors for each free parameter are listed in Table~\ref{tab:retrievals}.

\begin{deluxetable*}{lcccccccc}[ht!] \label{tab:retrievals}
\tablewidth{0pt}
\tablecaption{KELT-7b atmospheric retrieval priors and results.}
\tablehead{
\multicolumn{1}{c}{\hspace{-8em} \bfseries Retrieval Code} & \phm{-} &
\multicolumn{3}{c}{\bfseries \textsc{NEMESISPY}} & \phm{-} & \multicolumn{3}{c}{\bfseries POSEIDON}\\ \cmidrule{1-1} \cmidrule{3-5} \cmidrule{7-9}
Parameter & & Prior & No TESS & With TESS & \phm{-} & Prior & No TESS & With TESS
}
\startdata \\[-8pt]
\textbf{Composition} \\ 
\hspace{0.5em} log(H$_2$O) & & $\mathcal{U}$(-12,-1) & -5.28$_{-4.23}^{+3.09}$ & -3.27$_{-4.67}^{+1.22}$ & & $\mathcal{U}$(-12, -0.3) & $-6.81^{+3.01}_{-3.11}$ & $-6.54^{+3.12}_{-3.27}$ \\
\hspace{0.5em} log(CO$_2$) & & $\mathcal{U}$(-12,-1) & -2.95$_{-1.23}^{+0.79}$ & -3.11$_{-1.12}^{+0.79}$ & & $\mathcal{U}$(-12, -0.3) & $-4.12^{+1.13}_{-2.08}$ & $-4.23^{+1.08}_{-1.47}$ \\
\hspace{0.5em} log(CO) & & $\mathcal{U}$(-12,-1) & -6.48$_{-3.45}^{+2.98}$ & -6.82$_{-3.00}^{+3.10}$ & & $\mathcal{U}$(-12, -0.3) & $-6.81^{+3.11}_{-3.17}$ & $-6.92^{+3.13}_{-3.13}$ \\
\hspace{0.5em} log(TiO) & & $\mathcal{U}$(-12,-1) & -8.61$_{-2.00}^{+2.02}$ & -8.91$_{-1.90}^{+1.72}$ & & $\mathcal{U}$(-12, -0.3) & $-8.34^{+2.40}_{-2.23}$ & $-7.80^{+2.33}_{-2.55}$ \\
\hspace{0.5em} log(VO) & & $\mathcal{U}$(-12,-1) & -8.18$_{-2.33}^{+2.26}$ & -9.13$_{-1.75}^{+1.88}$ & & $\mathcal{U}$(-12, -0.3) & $-8.08^{+2.50}_{-2.39}$ & $-8.15^{+2.45}_{-2.33}$ \\
\hspace{0.5em} log(AlO) & & $\mathcal{U}$(-12,-1) & -8.26$_{-2.22}^{+2.45}$ & -8.61$_{-2.03}^{+2.23}$ & & $\mathcal{U}$(-12, -0.3) & $-8.00^{+2.44}_{-2.40}$ & $-7.97^{+2.37}_{-2.39}$ \\
\hspace{0.5em} log(SiO) & & $\mathcal{U}$(-12,-1) & -7.40$_{-2.79}^{+2.67}$ & -7.93$_{-2.35}^{+2.30}$ & & $\mathcal{U}$(-12, -0.3) & $-7.46^{+2.95}_{-2.78}$ & $-7.37^{+2.91}_{-2.85 }$ \\
\hspace{0.5em} log(FeH) & & $\mathcal{U}$(-12,-1) & -6.99$_{-2.91}^{+2.84}$ & -8.10$_{-2.36}^{+2.48}$ & & $\mathcal{U}$(-12, -0.3) & $-7.30^{+2.87}_{-2.90}$ & $-7.43^{+2.77}_{-2.76}$ \\
\hspace{0.5em} log(SH) & & $\mathcal{U}$(-12,-1) & -6.91$_{-3.05}^{+3.23}$ & -6.64$_{-3.21}^{+3.23}$ & & $\mathcal{U}$(-12, -0.3) & $-7.22^{+2.98}_{-2.92}$ & $-7.06^{+2.93}_{-3.01}$ \\
\hspace{0.5em} log(H$^-$) & & $\mathcal{U}$(-13,-2) & -4.94$_{-1.10}^{+0.89}$ & -6.20$_{-1.02}^{+0.82}$ & & $\mathcal{U}$(-12, -0.3) & $-5.32^{+0.75}_{-0.94}$ & $-5.79^{+0.83}_{-0.71}$ \\[5pt]
\textbf{Pressure-Temperature Profile} \\
\hspace{0.5em} $\alpha_1$ & & $\mathcal{U}$(0.02, 2.0) & $1.1_{-0.5}^{+0.5}$ & $1.1_{-0.5}^{+0.6}$ & & $\mathcal{U}$(0.02, 2.0) & $1.1^{+0.6}_{-0.6}$ & $1.1^{+0.6}_{-0.6}$ \\
\hspace{0.5em} $\alpha_2$ & & $\mathcal{U}$(0.02, 2.0) & $1.1_{-0.5}^{+0.5}$ & $1.1_{-0.6}^{+0.6}$ & & $\mathcal{U}$(0.02, 2.0) & $1.1^{+0.6}_{-0.6}$ & $1.1^{+0.6}_{-0.6}$ \\
\hspace{0.5em} log($P_1$) & & $\mathcal{U}$(-9, 2)  & $-3.2_{-3.0}^{+3.1}$ & $-3.1_{-3.0}^{+3.2}$ & & $\mathcal{U}$(-7, 2) & $-3.5^{+2.3}_{-2.1}$ & $-3.45^{+2.29}_{-2.16}$ \\
\hspace{0.5em} log($P_2$) & & $\mathcal{U}$(-9, 2)  & $-3.3_{-3.2}^{+3.2}$ & $-3.9_{-3.0}^{+3.5}$ & & $\mathcal{U}$(-7, 2) & $-3.4^{+2.2}_{-2.2}$ & $-3.33^{+2.27}_{-2.20}$ \\
\hspace{0.5em} log($P_3$) & & $\mathcal{U}$(max(log($P_1$,$P_2$)), 2) & $0.56_{-2.00}^{+1.08}$ & $0.51_{-2.12}^{+1.15}$ & & $\mathcal{U}$(-2, 2) & $0.45^{+1.01}_{-1.29}$ & $0.49^{+0.97}_{-1.24}$ \\
\hspace{0.5em} $T_{\mathrm{T.O.A.}}$(K) &  & $\mathcal{U}$(500, 3000)  & $1838_{-367}^{+385}$ & $1594_{-358}^{+372}$ & & --- & --- & --- \\
\hspace{0.5em} $T_{\mathrm{10 \, mbar}}$(K) & & --- & --- & --- & & $\mathcal{U}$(400, 3000) & $1891^{+516}_{-384}$ & $1784^{+415}_{-381}$ \\[5pt]
\textbf{Aerosol Properties} \\
\hspace{0.5em} log($P_{\mathrm{cloud}}$/bar) & & $\mathcal{U}$(-6, 1) & -4.0$_{-1.17}^{+2.15}$ & -2.4$_{-1.6}^{+1.9}$ & & $\mathcal{U}$(-7, 2) & $-3.0^{+2.9}_{-1.7}$ & $-2.4^{+2.6}_{-1.6}$ \\
\hspace{0.5em} $\alpha$ & & $\mathcal{U}$(0, 14) & 7.1 $_{-4.1}^{+4.1}$ & 7.0$_{-4.1}^{+3.9}$ & & --- & --- & --- \\
\hspace{0.5em} $\gamma$ & & --- & --- & --- & & $\mathcal{U}$(-20, 2) & $-11.1^{+5.8}_{-5.2}$ & $-11.2^{+4.5}_{-4.7}$ \\
\hspace{0.5em} log(opacity) & & $\mathcal{U}$(-10, 10) & -1.1$_{-5.0}^{+4.8}$ & -2.3$_{-4.7}^{+5.1}$ & & --- & --- & --- \\
\hspace{0.5em} log($a$) & & --- & --- & --- & & $\mathcal{U}$(-4, 8) & $4.2^{+2.5}_{-5.6}$ & $6.1^{+1.1}_{-6.4}$ \\[5pt]
\textbf{Other Planet Properties} \\
\hspace{0.5em} log($P_{\mathrm{ref}}$/bar) & & $\mathcal{U}$(-9, 2) & -7.27$_{-0.83}^{+1.05}$ & -7.80$_{-0.70}^{+0.97}$ & & --- & --- & --- \\
\hspace{0.5em} $R_{\mathrm{p, \, ref}}$(R$_{\mathrm{J}}$) & & --- & ---& --- & & $\mathcal{U}$(1.36, 1.84) & $1.48^{+0.02}_{-0.03}$ & $1.48^{+0.02}_{-0.02}$ \\[5pt]
\textbf{Stellar Properties} \\
\hspace{0.5em} $f_{\mathrm{het}}$ & & $\mathcal{U}$(0, 0.2) & 0.09$_{-0.04}^{+0.05}$ & 0.05$_{-0.03}^{+0.05}$ & & $\mathcal{U}$(0, 0.5) & $0.09^{+0.08}_{-0.05}$ & $0.06^{+0.07}_{-0.04}$ \\
\hspace{0.5em} $T_{\mathrm{het}}$(K) & & --- & --- & --- & & $\mathcal{U}$(4738, 8122) & $7064^{+237}_{-155}$ & $7052^{+289}_{-398}$ \\
\hspace{0.5em} $\Delta T_{\mathrm{het}}$(K) & & $\mathcal{U}$(-750, 750) & 333$_{-126}^{+161}$ & 103$_{-125}^{+190}$ & & --- & --- & --- \\
\hspace{0.5em} $T_{\mathrm{phot}}$(K) & & $\mathcal{N}$(6848, 200) & 6848$_{-139}^{+159}$ & 6807$_{-161}^{+173}$ & & $\mathcal{N}$(6768, 7) & $6768^{+6}_{-6}$ & $6768^{+6}_{-6}$ \\[5pt]
\textbf{Offsets} \\
\hspace{0.5em} WFC3-UVIS (ppm) & & $\mathcal{N}$(0, 70) & -48$_{-40}^{+38}$ & -191$_{-21}^{+22}$ & & $\mathcal{U}$(-500, 500) & $109^{+58}_{-54}$ & $204^{+33}_{-30}$ \\
\hspace{0.5em} Spitzer-UVIS (ppm) & & $\mathcal{N}$(0, 60) & -78$_{-65}^{+57}$ & -128$_{-60}^{+53}$ & & $\mathcal{U}$(-500, 500) & $283^{+122}_{-118}$ & $352^{+89}_{-100}$ \\[3pt]
\enddata
\tablecomments{$R_{\mathrm{p, \, ref}}$ is defined at 10~bar for \textsc{POSEIDON}. `---' denotes a parameter used in one retrieval code but not in the other (e.g. different P-T and aerosol parameterizations). The top of atmosphere pressure in \textsc{NEMESISPY} is 10$^{-9}$ atm.}
\end{deluxetable*}

\subsection{Retrieval Results: HST UVIS + IR + \text{Spitzer}}
\label{sec:retrievalresults}

Our retrieval analysis finds that KELT-7b's transmission spectrum is best fit by H$^-$ bound-free absorption alongside contamination from bright stellar inhomogeneities. We do not confidently detect any other gases in the atmosphere. Our retrievals indicate the possible presence of CO$_2$, but since this inference is driven entirely by the two broadband Spitzer IRAC points the reliability of this result cannot be definitively established with the present data. Future spectroscopic observations with JWST will be able to verify or refute the presence of CO$_2$. The median spectra and $\pm1\sigma$ and $\pm2\sigma$ confidence regions retrieved with \textsc{POSEIDON} and \textsc{NEMESISPY} are shown in the left and right panels of Figure~\ref{fig:ret_spectra}, respectively. Figure~\ref{fig:corners} highlights a subset of the posterior distribution for well-constrained model parameters from both retrieval codes. The full retrieval results are presented in Table~\ref{tab:retrievals}.

The lower transit depths in the near-UV are caused by bright stellar inhomogeneities such as hot stellar active regions. The additional flux from these unocculted regions of the stellar disk causes a negative slope towards shorter wavelengths, which cannot be explained by atmospheric features \citep[e.g.][]{Rackham2018,rathcke2021}. Both retrieval codes find hot active regions covering $\approx$ 10\% of the stellar surface ($8^{+5}_{-4}$\% for \textsc{NEMESISPY}, $9^{+8}_{-4}$\% for \textsc{POSEIDON}) at $\approx$ 300\,K warmer than the stellar photosphere ($344^{+169}_{-135}$\,K for \textsc{NEMESISPY}, $296^{+235}_{-156}$\,K for \textsc{POSEIDON}). We note that both codes yield consistent active region properties despite different priors on the stellar photosphere temperature (\textsc{POSEIDON} used a 7\,K standard deviation based on \citet{Stassun2017}, while \textsc{NEMESISPY} used a more agnostic 200\,K, which explains the sharp $T_{\rm{phot}}$ posteriors for \textsc{POSEIDON} in Figure~\ref{fig:corners}). To evaluate the significance of the stellar inhomogeneities we ran an additional \textsc{NEMESISPY} retrieval without including the active regions as free parameters. We then compute the detection significance  as the difference in Bayesian evidence between the retrievals with and without stellar activity, and transform the corresponding Bayes factor to standard deviation via the conversion presented in \cite{trotta2008}. As a result, we obtain a red Bayes factor ($\mathcal{B}$) of 10 for the presence of stellar inhomogeneities (2.7$\sigma$ significance).

Our retrieval results yield strong detections of H$^-$ bound-free opacity shaping KELT-7b's transmission spectrum. We compute the detection significance of H$^-$ following the same procedure as previously described for the stellar inhomogeneities, obtaining $\mathcal{B} = 17,000$ (4.8$\sigma$) with \textsc{NEMESISPY} and $\mathcal{B} = 6,000$ (4.6$\sigma$) with \textsc{POSEIDON}. The retrieved H$^{-}$ abundances exceed 1\,ppm, with consistent abundances between the two codes of $\log \rm{H^{-}} = -4.94_{-1.10}^{+0.89}$ (\textsc{NEMESISPY}) and $\log \rm{H^{-}} = -5.32^{+0.75}_{-0.94}$ (\textsc{POSEIDON}). 

We do not detect H$_2$O in KELT-7b's atmosphere. Though neither retrieval detects H$_2$O, the upper limit on the allowed H$_2$O abundance is relatively weak (e.g. $\log \rm{H_2 O} < -2.37$ to 2$\sigma$ with \textsc{POSEIDON}). Therefore, even a solar abundance of H$_2$O ($\log \rm{H_2 O} \approx -3.3$) would not result in a H$_2$O detection in these observations. This non-detection of H$_2$O results from the H$^{-}$ absorption overwhelming the spectral shape in the infrared wavelengths probed by WFC3/IR G141. 
The solutions for both retrieval codes allow a high-altitude grey cloud deck, but neither retrieval code unambiguously detects a cloud deck. Figure~\ref{fig:corners} shows a modal peak around $\log P_{\rm{cloud}} \sim$ 30\,$\upmu$bar for both \textsc{POSEIDON} and \textsc{NEMESISPY}, but the posteriors display a long tail towards higher cloud pressures without the  distinct upper bound on $\log P_{\rm{cloud}}$ that would be seen if the data required a cloud deck. The broad uncertainty on the presence of clouds ($\log (P_{\rm{cloud}} / \rm{bar}) = 4.0^{+2.15}_{-1.17}$ for \textsc{NEMESISPY} and $\log (P_{\rm{cloud}} / \rm{bar}) = -2.96^{+2.83}_{-1.68}$ \textsc{POSEIDON}) is due to the degeneracy between the cloud top pressure and the H$^-$ abundance (see Figure~\ref{fig:corners}), since both clouds and H$^-$ can produce a relatively featureless near-UV transmission spectrum. However, despite this degeneracy, a cloud deck alone cannot account for the shape of KELT-7b's WFC3 G141 spectrum. While the present data are inconclusive on the presence of high-altitude clouds, our retrievals nevertheless require the presence of H$^{-}$ even when clouds are considered. As usual with Hubble transit observations, we are unable to draw any strong conclusions about the temperature structure of KELT-7b; the retrieved pressure-temperature profile with both \textsc{POSEIDON} and \textsc{NEMESISPY} is essentially an isothermal profile centered at the corresponding maximum likelihood temperature, with fairly broad uncertainties.

Additionally, we evaluate the predictions from equilibrium chemistry by running forward models with \textsc{NEMESISPY} and Fastchem \citep{stock2018, stock2022}. We assume an isothermal T-P profile at the maximum likelihood temperature retrieved from the \textsc{NEMESISPY} free retrievals, and add the effect from stellar activity and clouds. The forward models are not able to appropriately fit the observed spectrum, mainly due to the inability to produce sufficient H$^{-}$ under equilibrium conditions. In particular, we obtain deep atmosphere equilibrium abundances of the order of $10^{-12}$ for H$^{-}$; several orders of magnitude below the abundances obtained in the free retrievals (log H$^-$ $\approx$ -5). This essentially indicates that the H$^-$ abundance is being enhanced by disequilibrium processes, and more particularly photochemistry, taking place in the upper atmosphere of KELT-7b. For the conditions relevant to KELT-7b, the high H$^{-}$ abundance is likely explained by substantial dissociation of H$_2$ to produce atomic H, the precursor for H$^{-}$ production, alongside sufficient ionization of other chemical species to provide free electrons \citep{lewis2020}. Given that at these temperatures enough H$_2$ is already dissociated, it is likely the insufficient ionization, and therefore the lack of free electrons, what precludes the creation of H$^-$ under equilibrium conditions. Furthermore, we observe that CO dominates under equilibrium conditions, as expected from KELT-7b's high equilibrium temperature. For a C/O ratio of 0.5 we obtain a CO$_2$ abundance of $10\times$ solar, which is significantly less than the abundance obtained with the free retrievals. We however reiterate that the conclusions regarding CO and CO$_2$ are solely driven by the Spitzer data points, which are unable to definitively resolve between different absorbers.

\subsection{Retrieval Results: Inclusion of TESS data}

\begin{figure*}
    \centering
    \includegraphics[width = \textwidth]{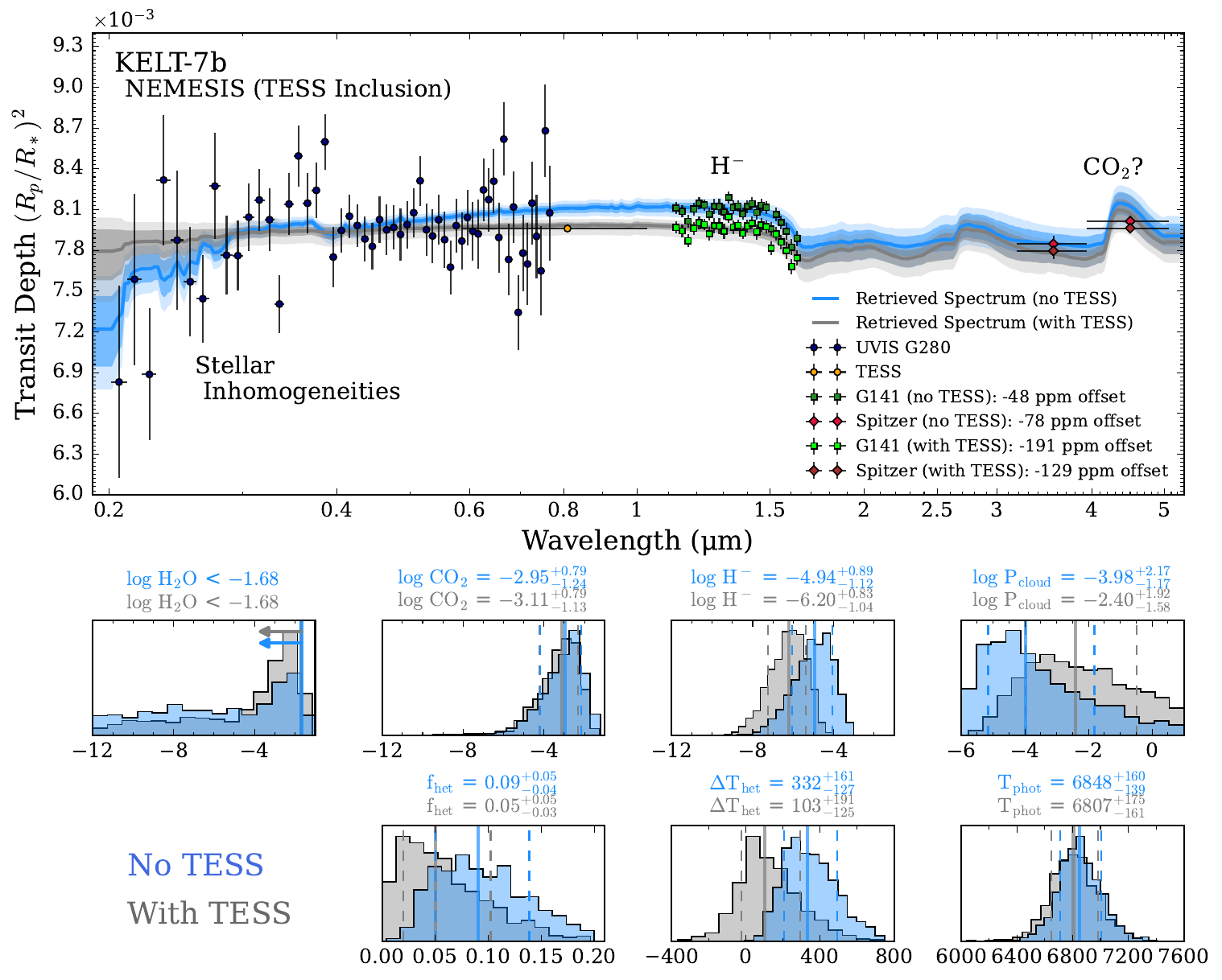}
    \caption{Atmospheric retrievals of KELT-7b's transmission spectrum conducted with \textsc{NEMESISPY} including the TESS data point (gray). For comparison, the results without the TESS data are also plotted (blue). The top panel shows the median retrieved spectrum. The histograms on the lower panels show the posterior probability distributions of a subset of the parameters retrieved with \textsc{NEMESISPY}.}
    \label{fig:ret_tess}
\end{figure*}

The NASA TESS mission \citep{ricker2014} observed nine transits of KELT-7b with 2-minute cadence during sector 19, observed between November 28 and December 23 2019. Those data were analyzed by \cite{pluriel2020}, who reported a transit depth of 0.7961 $\pm$ 0.0018 $\%$ over the integrated TESS photometric passband of 0.4 $\mu$m, centered at 0.8 $\mu$m.
We performed additional \textsc{NEMESISPY} and \textsc{POSEIDON} retrievals including the TESS photometric datapoint from \cite{pluriel2020}. We do not fit for an additional offset between the TESS data point and the UVIS data, since allowing an offset for a single point would render its inclusion in the retrieval effectively meaningless; instead, we anchor it to the UVIS data, with which it overlaps, and retrieve for offsets on WFC3 and Spitzer relative to UVIS as before.  Figure~\ref{fig:ret_tess} shows the retrieved median spectrum and the posterior distribution of a subset of parameters obtained when including the TESS data point. The full retrieval results are shown in Table~\ref{tab:retrievals}. The presence of H$^-$ is not significantly affected by the addition of the TESS data, as we retrieve similar abundances to the ones obtained before ($\log \rm{H^{-}} = -6.20^{+0.82}_{-1.02}$ with \textsc{NEMESISPY}; $\log \rm{H^{-}} = -5.50^{+0.63}_{-0.93}$ with \textsc{POSEIDON}). Including the TESS data point, however,  
increases the retrieved abundance of H$_2$O (although the long tail towards lower values remains).

The addition of the TESS data point also lessens the effect of stellar contamination, reducing the bright inhomogeneity coverage fraction to $\approx 5\%$ and the temperature difference between the stellar inhomogeneities and the photosphere to $\approx$ 150 K.
The TESS data point has such a significant influence because the error bars on this point are the smallest of any in our dataset. However, we believe that the uncertainty as to whether the TESS baseline is indeed consistent with that of the UVIS data points exceeds the quoted uncertainty. For example, for the retrievals not including TESS we retrieve offsets of -50 ppm and -87 ppm for the WFC3-IR and Spitzer spectra, respectively, while the uncertainty on the TESS point is 18 ppm. These retrieval offsets increase substantially when TESS is included in the retrieval analysis. For this reason, we consider that the TESS data point does not add any meaningful information to the retrieval, since we have no way of accounting for a likely baseline offset, and so our preferred solution excludes this point. 

\subsection{Stellar inhomogeneities and faculae} \label{sec:faculae}

Both retrieval codes find that the lower transit depths observed in the near-UV are better fit when including the effect of bright inhomogeneities, such as hot active regions, on the surface of the star. As described in the previous sections, the retrievals fit for the coverage and temperature of regions colder or hotter than the photosphere. While in previous studies the term ``stellar faculae" has been typically used to refer to such inhomogeneities in the stellar surface, the presence of stellar faculae is the result of a well-known physical process that can only take place under certain conditions. Solar faculae occur when the convective motion of the stellar envelope transports small magnetic flux tubes into intergranular lanes, where the field is intensified by convective collapse \citep{parker1978, spruit1979, solanki1993small, johnson2021}. As a result, the increased radiative heating from the hot walls produces regions brighter than the surrounding photosphere, commonly known as faculae \citep{parker1978, spruit1979}. KELT-7 is a fast rotating star ($\sim 73\pm0.5$ km/s) with a mass of $\sim1.5M_{\odot}$ \citep{bieryla2015}. This indicates that the star must have a radiative envelope with no strong magnetic activity, hence making the presence of stellar faculae in KELT-7 infeasible. We therefore avoid using the term stellar faculae throughout this paper, and instead attribute the observed spectrum to the presence of inhomogeneities in the stellar surface shaped by a different physical process. 

\begin{figure*}
    \centering
    \includegraphics[width = \textwidth]{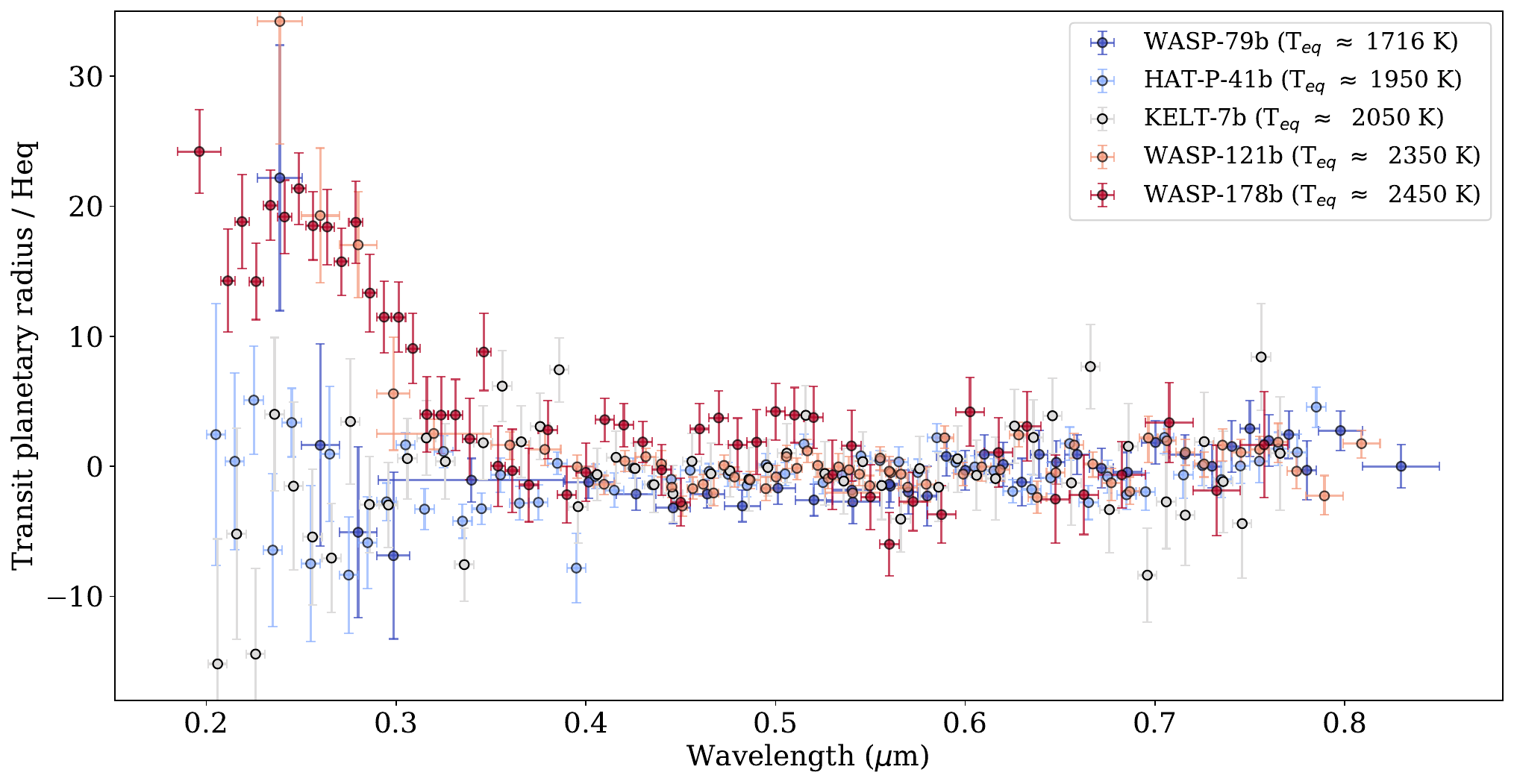}
    \caption{Transmission spectrum of KELT-7b compared to the WFC3/UVIS G280 UV-optical transmission spectra of WASP-178b \citep{lothringer2022} and HAT-P-41b \citep[]{wakeford2020}, and the STIS E230M, G430L and G750L transmission spectra of WASP-79b \citep{rathcke2021, gressier2023} and WASP-121b \citep{sing2019}.}
    \label{fig:comp}
\end{figure*}

\section{Comparison to other Hot Jupiters with UV-optical observations}
\label{sec:comp}

We compare our newly obtained WFC3/UVIS G280 transmission spectrum of KELT-7b, to the spectra of other hot Jupiters with published UV-optical observations. Those planets are HAT-P-41b \citep{wakeford2020} and WASP-178b \citep{lothringer2022}, observed with WFC3/UVIS G280, and WASP-121b \citep{evans2018, sing2019} and WASP-79b \citep{rathcke2021, gressier2023}, observed with HST's Space Telescope Imaging Spectrograph (STIS) E230M, G430L and G750L gratings. The spectra of all these planets, normalized by their atmospheric scale height, are shown in Figure~\ref{fig:comp}. 
The shape of these transmission spectra, specially towards the bluer end, reveal two clear regimes: on the one hand, observations of the ultra-hot Jupiters WASP-178b ($ \rm T_{eq} = 2450K$) and WASP-121b ($ \rm T_{eq} = 2350K$) show strong absorption in the UV, with features of 10-30 atmospheric scale heights in depth. However, with slightly lower equilibrium temperatures, HAT-P-41b ($ \rm T_{eq} = 1937K$), KELT-7b ($ \rm T_{eq} = 2090K$), and WASP-79b ($ \rm T_{eq} = 1716K$), show a relatively flat spectrum with no absorption in the UV, with the exception of one point at the bluer end of WASP-79b's spectrum. Atmospheric retrievals conducted on WASP-178b and WASP-121b attributed the UV absorption to the presence of gaseous refractory species such as Fe, Mg and SiO \citep{lothringer2022, sing2019}, all of which have significant opacity at NUV wavelengths. At lower temperatures, these gaseous elements are expected to condensate into silicate cloud particles like MgSiO$_3$ and Mg$_2$SiO$_4$ \citep{visscher2010}, causing the spectra to be partially or totally muted. The condensation temperature at which this clear-to-cloudy transition takes place is yet to be constrained. As pointed out in \cite{lothringer2022}, the spectra shown in Figure~\ref{fig:comp} suggest that this transition could take place somewhere in between 1950K and 2450K. If we were to assume that (at least part of) the flatness observed in the UV-optical spectrum of KELT-7b is attributed to the presence of a prominent cloud deck, our observations would suggest that this condensation would occur around $ \rm T_{eq} \approx 2100K$. However, while the solutions for both retrieval codes allow for a high-altitude grey cloud deck, the uncertainties are significant and therefore neither retrieval code can unambiguously detect a cloud deck (Section~\ref{sec:retrievalresults}).

The absence of spectral features observed in the three colder planets, however, can also be explained (in part) by the presence of a dominant UV-optical opacity source. As shown in Section~\ref{sec:retrievals}, our retrievals favour the presence of H$^-$, for which we detect relatively high abundances . In turn, our retrievals are not able to place strong constraints on the cloud top pressures. Similarly, retrievals performed on HAT-P-41b indicated that, while a significant cloud deck could be responsible for the observed spectrum, the latter was better explained by the presence of a strong chemical opacity source, most likely H$^-$, without the need for clouds in the observable atmosphere \citep{lewis2020}. Finally, \cite{rathcke2021} concluded that the atmosphere of WASP-79b was better explained by the presence of H$^-$ in combination with H$_{2}$O, with no detected cloud opacity. The degeneracy between cloud top pressure and H$^-$ abundance hinders our ability to precisely determine the contribution from each opacity source. However, in all three cases, clouds alone are not able to account for the observed spectrum. High abundances of H$^-$, on the other hand, appear to be the main contributor to the flatness observed in the spectra these hot Jupiters. Under this assumption, the cloudy-to-clear transition could potentially happen at lower temperatures, with the presence of H$^-$ being the main driver of the spectral shape observed in HAT-P-41b, KELT-7b and WASP-79b.

\section{Summary and Conclusions}
\label{sec:conclusions}

We presented new UV-optical observations of the hot Jupiter KELT-7b, obtained with HST's WFC3/UVIS G280 as part of the HUSTLE treasury program. We reduced the observations using two independent pipelines, \lluvia and \Hazelnut \citep{Boehm2024}, which showed good agreement between the derived spectra. We combined our observations with HST WFC3/IR G141 data and \textit{Spitzer} IRAC data from \cite{pluriel2020}  to create a $0.2 - 1.7 \micron$ transmission spectrum. We then performed atmospheric retrievals on this spectrum with the \textsc{POSEIDON} and \textsc{NEMESISPY} retrieval codes, in order to infer the properties of the planet and the host star. We also performed additional retrievals including the TESS photometric data point from \cite{pluriel2020}. Finally, we compared the spectrum of KELT-7b spectrum to those of other hot Jupiters with UV-optical observations, and discussed the implications of our results. We summarize our main findings here:

\begin{itemize}
    \item Our measured HST WFC3/UVIS G280 transmission spectrum is relatively flat, with no clear absorption features and a downward slope toward the UV starting at $\simeq 0.3\micron$.
    \item Both retrievals find that the combined HST WFC3/UVIS, WFC3/IR and Spitzer transmission spectrum is best fit by a planetary atmosphere dominated by H$^-$ absorption with contamination from stellar inhomogeneities. While we do not find clear evidence of H$_2$O, we point out that substantial amounts of H$_2$O are well within the uncertainty range of our retrieved spectra. We find tentative signs of CO$_2$. However, this is solely driven by the addition of the \textit{Spitzer} IRAC data points, which could also be explained, for instance, by the presence of CO. The solutions for both retrieval codes are consistent with grey clouds high in the atmosphere, but the degeneracy with the H$^-$ abundance causes the retrieved cloud top pressure to be relatively unconstrained.
    \item We detect  H$^-$ with high significance, with recovered abundances of log(H$^-$)   = -4.94$_{-1.10}^{+0.89}$ and log(H$^-$) = $-5.32^{+0.75}_{-0.94}$ for \textsc{NEMESISPY} and \textsc{POSEIDON} respectively. We add KELT-7b to the list of gas giants around F-stars with signs of H$^-$, which suggests that the increased UV flux from these stars might be responsible for an enhanced photoionisation of alkali metals, which would ultimately lead to the onset of observable H$^-$.
\end{itemize}

Future observations with JWST will help place constraints on the presence and abundance of H$^-$ and ${\rm CO_2}$ in KELT-7b. In particular, planned observations with NIRISS SOSS (GO 5924, PI: Sing) and NIRSpec G395H (GO 3838, PIs: Kirk \& Ahrer) will be able to probe the shape of KELT-7b's spectrum from 0.6 to 2.8 $\micron$, where the absorption from H$^-$ can be better identified, and from 2.87 to 5.14 $\micron$ , where the presence or absence of ${\rm CO_2}$ can be confirmed. Furthermore, observations within the HUSTLE program will allow us to obtain the UV-optical spectrum of numerous other targets spanning temperatures from 960K up to 2650K. As a result, we will be able to identify which other targets show clear signs of H$^-$ absorption, as well as constrain the temperature at which the clear to cloudy transition takes place.

\vspace{5mm}

This research is based on observations made with the NASA/ESA \textit{Hubble Space Telescope} obtained from the Space Telescope Science Institute, which is operated by the Association of Universities for Research in Astronomy, Inc., under NASA contract NAS 5–26555. These observations are associated with program HST-GO 17183, PI: H.R. Wakeford. This research has made use of the NASA Exoplanet Archive, which is operated by the California Institute of Technology, under contract with the National Aeronautics and Space Administration under the Exoplanet Exploration Program. 
C.G. acknowledges funding from La Caixa Fellowship and the Agency for Management of University and Research Grants from the Government of Catalonia (FI AGAUR).
R.J.M and S.E.M. are supported by NASA through the NASA Hubble Fellowship grant HST-HF2-51563 awarded by the Space Telescope Science Institute, which is operated by the Association of Universities for Research in Astronomy, Inc., for NASA, under contract NAS5-26555. 
J.K.B. is supported by a Science and Technology Facilities Council Ernest Rutherford Fellowship ST/T004479/1.
H.R.W. and D.G were funded by UK Research and Innovation (UKRI) framework under the UK government’s Horizon Europe funding guarantee for an ERC Starter Grant [grant number EP/Y006313/1]. L.A is supported by the Klarman Fellowship and acknowledges funding from the UKRI STFC Consolidated Grant ST/V000454/1. 
C.E.F. acknowledges funding from the University of Bristol School of Physics PhD Scholarship Fund. 


\vspace{5mm}
\facility{\\
    HST(WFC3)}

\software{\\ 
    astropy \citep{astropy:2013, astropy:2018, astropy:2022} \\
    scipy \citep{scipy2020} \\
    ExoTiC-LD \citep{grant_2022} \\
    ExoTiC-ISM \citep{Exoticism} \\
    batman \citep{kreidberg2015} \\
    emcee \citep{foreman2013} \\
    \href{https://github.com/MartianColonist/POSEIDON}{\textsc{POSEIDON}} \citep{MacDonald2017a,MacDonald2023} \\
    \href{https://github.com/Jingxuan97/nemesispy}{\textsc{NEMESISPY}} \citep{irwin2008,yang2024}}

\begin{deluxetable*}{ccccccc}
\tabletypesize{\footnotesize}
\renewcommand{\arraystretch}{0.95}
\setlength{\tabcolsep}{10pt}
\tablecaption{KELT-7b transmission spectrum results. The transit depth $(R_{p}/R_{*})^2$ and the corresponding uncertainty are given for the combined, +1 order and -1 order spectra.}
\label{tab:transmspec}
\tablehead{
\\[-13pt]
\colhead{} & \multicolumn{2}{c}{Combined} & \multicolumn{2}{c}{+1 Order} & \multicolumn{2}{c}{-1 Order}\\[-2pt] \cmidrule(lr){2-3} \cmidrule(lr){4-5} \cmidrule(lr){6-7} \\[-15pt]
\colhead{Wavelength} & \colhead{Transit depth} &  \colhead{Uncertainty} &  \colhead{Transit depth} & \colhead{Uncertainty} & \colhead{Transit depth} & \colhead{Uncertainty} \\[-5pt]
\colhead{(nm)} & \colhead{($\%$)} & \colhead{($\%$)} & \colhead{($\%$)} & \colhead{($\%$)} & \colhead{($\%$)} & \colhead{($\%$)} 
}
\startdata
206 & 0.68310 & 0.07080 & 0.68310 & 0.07080 & -       & -       \\
216 & 0.75870 & 0.06300 & 0.75870 & 0.06300 & -       & -       \\
226 & 0.68870 & 0.04860 & 0.68870 & 0.04860 & -       & -       \\
236 & 0.83180 & 0.04800 & 0.86000 & 0.05340 & 0.71330 & 0.10950 \\
246 & 0.78750 & 0.05100 & 0.75670 & 0.05910 & 0.87760 & 0.10090 \\
256 & 0.75690 & 0.04040 & 0.75930 & 0.04520 & 0.74710 & 0.09040 \\
266 & 0.74440 & 0.03220 & 0.73560 & 0.03650 & 0.77560 & 0.06840 \\
276 & 0.82730 & 0.03900 & 0.84450 & 0.04330 & 0.75370 & 0.08940 \\
286 & 0.77640 & 0.02900 & 0.76890 & 0.03250 & 0.80590 & 0.06420 \\
296 & 0.77610 & 0.02570 & 0.79120 & 0.02980 & 0.73190 & 0.05100 \\
306 & 0.80440 & 0.02470 & 0.79680 & 0.02910 & 0.82360 & 0.04640 \\
316 & 0.81710 & 0.02300 & 0.82220 & 0.02660 & 0.80140 & 0.04640 \\
326 & 0.80260 & 0.02300 & 0.81850 & 0.02790 & 0.76880 & 0.04070 \\
336 & 0.74050 & 0.02140 & 0.74920 & 0.02480 & 0.71570 & 0.04190 \\
346 & 0.81410 & 0.02280 & 0.81440 & 0.02690 & 0.81350 & 0.04320 \\
356 & 0.84950 & 0.02240 & 0.85440 & 0.02620 & 0.83620 & 0.04300 \\
366 & 0.81470 & 0.02210 & 0.82410 & 0.02570 & 0.78780 & 0.04340 \\
376 & 0.82420 & 0.02060 & 0.83100 & 0.02370 & 0.80340 & 0.04170 \\
386 & 0.85990 & 0.02040 & 0.84960 & 0.02570 & 0.87770 & 0.03360 \\
396 & 0.77510 & 0.02210 & 0.78960 & 0.02840 & 0.75310 & 0.03500 \\
406 & 0.79460 & 0.01610 & 0.80330 & 0.01930 & 0.77430 & 0.02930 \\
416 & 0.80510 & 0.01600 & 0.79550 & 0.01870 & 0.83030 & 0.03050 \\
426 & 0.79830 & 0.01560 & 0.80220 & 0.01870 & 0.78950 & 0.02810 \\
436 & 0.78840 & 0.01670 & 0.76730 & 0.01970 & 0.84210 & 0.03140 \\
446 & 0.78290 & 0.01720 & 0.76450 & 0.02190 & 0.81270 & 0.02780 \\
456 & 0.80270 & 0.01700 & 0.81540 & 0.02140 & 0.78120 & 0.02790 \\
466 & 0.79510 & 0.01860 & 0.79790 & 0.02220 & 0.78840 & 0.03440 \\
476 & 0.79690 & 0.01610 & 0.78600 & 0.01980 & 0.81820 & 0.02770 \\
486 & 0.79170 & 0.01940 & 0.79410 & 0.02430 & 0.78770 & 0.03210 \\
496 & 0.79890 & 0.01740 & 0.79400 & 0.02140 & 0.80840 & 0.02970 \\
506 & 0.80770 & 0.01810 & 0.78080 & 0.02300 & 0.85160 & 0.02930 \\
516 & 0.83120 & 0.01830 & 0.83870 & 0.02160 & 0.81270 & 0.03410 \\
526 & 0.79520 & 0.01940 & 0.79490 & 0.02410 & 0.79560 & 0.03260 \\
536 & 0.79060 & 0.01990 & 0.77700 & 0.02370 & 0.82330 & 0.03670 \\
546 & 0.80250 & 0.01830 & 0.81680 & 0.02260 & 0.77510 & 0.03120 \\
556 & 0.78790 & 0.02070 & 0.80840 & 0.02670 & 0.75720 & 0.03270 \\
566 & 0.76760 & 0.01980 & 0.75270 & 0.02380 & 0.80130 & 0.03580 \\
576 & 0.79820 & 0.01960 & 0.78620 & 0.02450 & 0.81960 & 0.03270 \\
586 & 0.78690 & 0.02090 & 0.76270 & 0.02510 & 0.84080 & 0.03750 \\
596 & 0.80420 & 0.02040 & 0.80560 & 0.02500 & 0.80120 & 0.03540 \\
606 & 0.79410 & 0.02140 & 0.80790 & 0.02630 & 0.76680 & 0.03700 \\
616 & 0.79210 & 0.02540 & 0.75890 & 0.03180 & 0.84970 & 0.04200 \\
626 & 0.82450 & 0.02300 & 0.78950 & 0.02980 & 0.87660 & 0.03630 \\
636 & 0.81740 & 0.02340 & 0.79630 & 0.02780 & 0.86860 & 0.04340 \\
646 & 0.83090 & 0.02360 & 0.85170 & 0.02870 & 0.78730 & 0.04140 \\
656 & 0.78950 & 0.02570 & 0.76480 & 0.03360 & 0.82460 & 0.04000 \\
666 & 0.86190 & 0.02690 & 0.86170 & 0.03340 & 0.86240 & 0.04550 \\
676 & 0.77330 & 0.02610 & 0.77850 & 0.03160 & 0.76220 & 0.04640 \\
686 & 0.81190 & 0.02650 & 0.83000 & 0.03300 & 0.77870 & 0.04460 \\
696 & 0.73430 & 0.02760 & 0.74850 & 0.03460 & 0.70970 & 0.04560 \\
706 & 0.77800 & 0.02830 & 0.75800 & 0.03490 & 0.81650 & 0.04840 \\
716 & 0.77000 & 0.03020 & 0.78030 & 0.04020 & 0.75650 & 0.04570 \\
726 & 0.81480 & 0.03050 & 0.79330 & 0.03750 & 0.85690 & 0.05240 \\
736 & 0.79030 & 0.03090 & 0.80320 & 0.03970 & 0.77020 & 0.04930 \\
746 & 0.76490 & 0.03270 & 0.76120 & 0.03970 & 0.77280 & 0.05740 \\
756 & 0.86800 & 0.03390 & 0.84560 & 0.04220 & 0.90880 & 0.05700 \\
766 & 0.80760 & 0.03490 & 0.81020 & 0.04510 & 0.80370 & 0.05510
\enddata

\end{deluxetable*}

\appendix

\section{Retrievals without UVIS}

\begin{figure*}
    \centering
    \includegraphics[width = \textwidth]{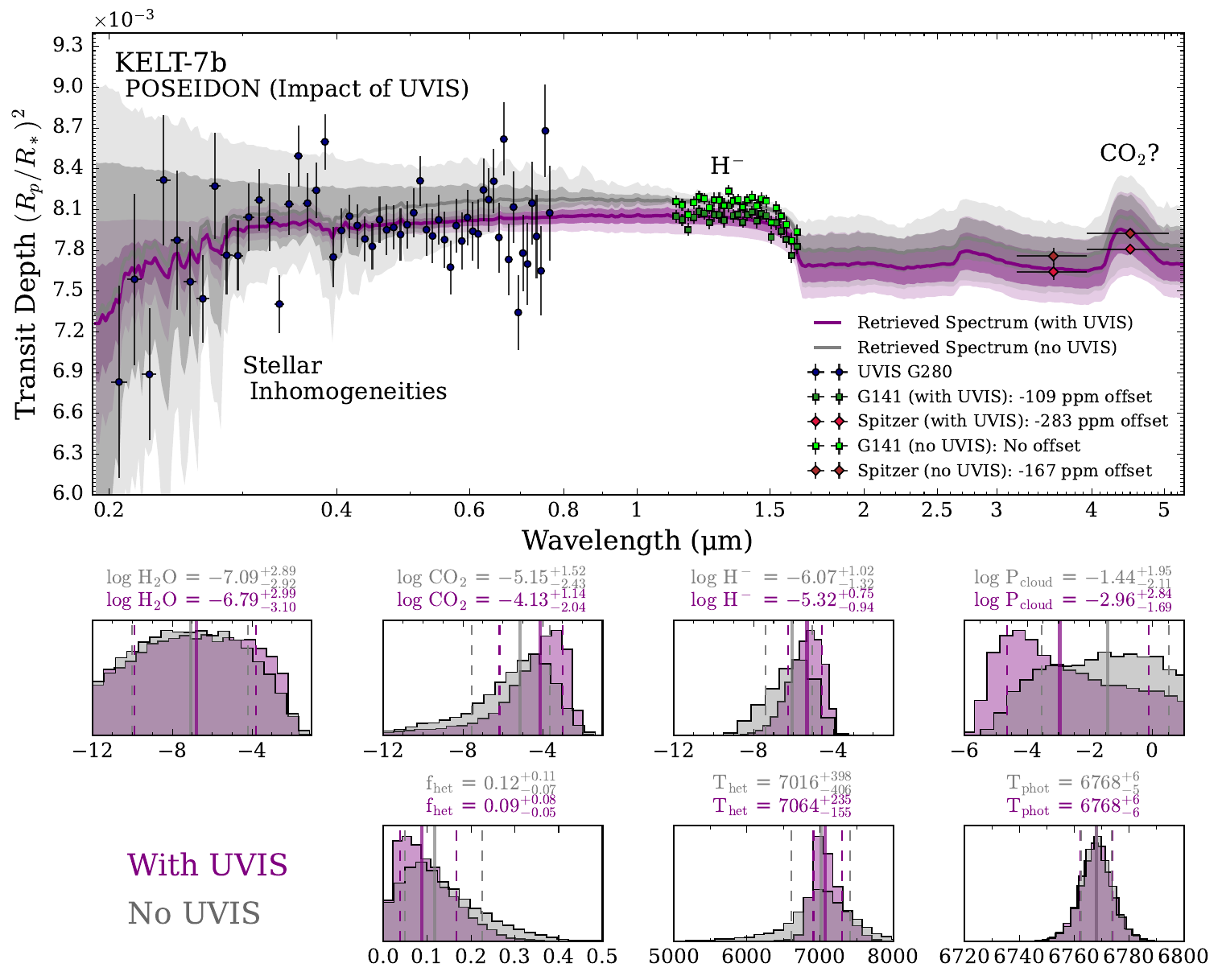}
    \caption{Atmospheric retrievals of KELT-7b's transmission spectrum conducted with \textsc{POSEIDON} without including the WFC3/UVIS G280 data (gray). For comparison, the results with the UVIS data are also plotted (purple). The top panel shows the median retrieved spectrum. The histograms on the lower panels show the posterior probability distributions of a subset of the parameters retrieved with \textsc{POSEIDON}.}
    \label{fig:ret_not_uvis}
\end{figure*}

To evaluate and quantify the information provided by the new UV-optical observations presented in this study, we have conducted additional retrievals without including the WFC3/UVIS G280 data, and  only using the WFC3/IR G141 data from \cite{pluriel2020} and the Spitzer data from \cite{baxter2021}. Figure~\ref{fig:ret_not_uvis} shows the histograms corresponding to the posterior probability distributions of a subset of the parameters retrieved with and without the UVIS data, shown in purple and gray, respectively. While \ce{H-} is still detected with the infrared data alone, the inclusion of the UVIS data allows us to place a tighter constraint on the corresponding \ce{H-} abundance ($\log \rm{H^{-}}$  = $-5.32^{+0.75}_{-0.94}$). The broad posteriors obtained for the temperature of the stellar heterogeneities $T_{\mathrm{het}}$ indicate that, without the UVIS data, the retrievals are unable to distinguish between the presence of colder (spots) or hotter active regions in the stellar photosphere. In contrast, when including the UVIS data, $T_{\mathrm{het}}$ is significantly better constrained which points towards the presence of bright stellar inhomogeneities. Finally, while the infrared data alone are unable to place constraints on the cloud top pressure, the inclusion of the UVIS data indicates that a lower cloud top pressure is slightly favored.

\section{WFC3/UVIS G280 Light curves} \label{sec:anexlcs}

\begin{figure*}
    \centering
    \includegraphics[width = \textwidth]{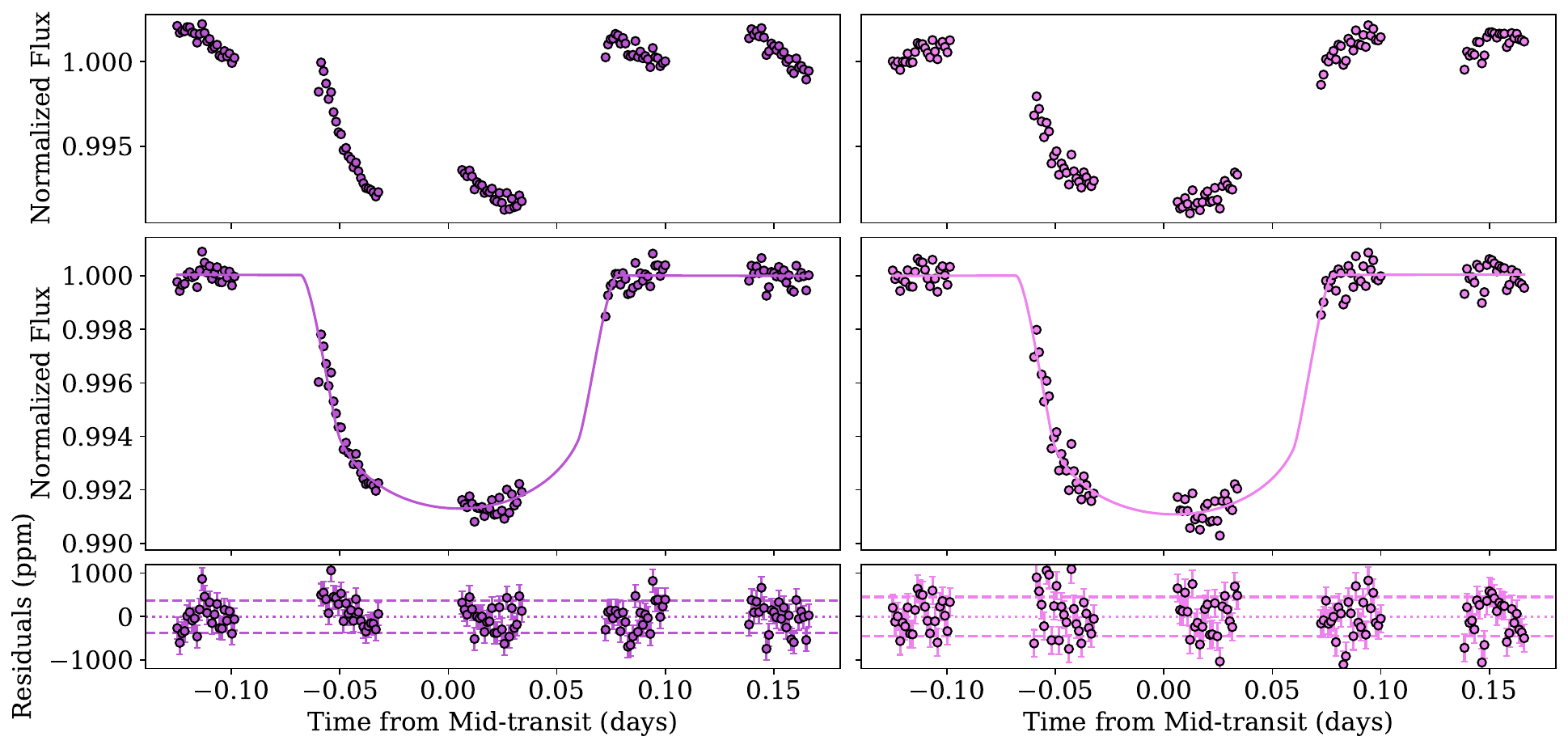}
    \caption{Top: Raw +1 (left) and -1 (right) order white-light curve obtained with \Hazelnut. Bottom: Systematics corrected +1 (left) and -1 (right) order white-light curve. The narrow panel below shows the transit fit residuals, with the dashed lines indicating the standard deviation of the residuals.}
    \label{fig:wlc_abby}
\end{figure*}

\begin{figure*}
    \centering
    \includegraphics[width = \textwidth]{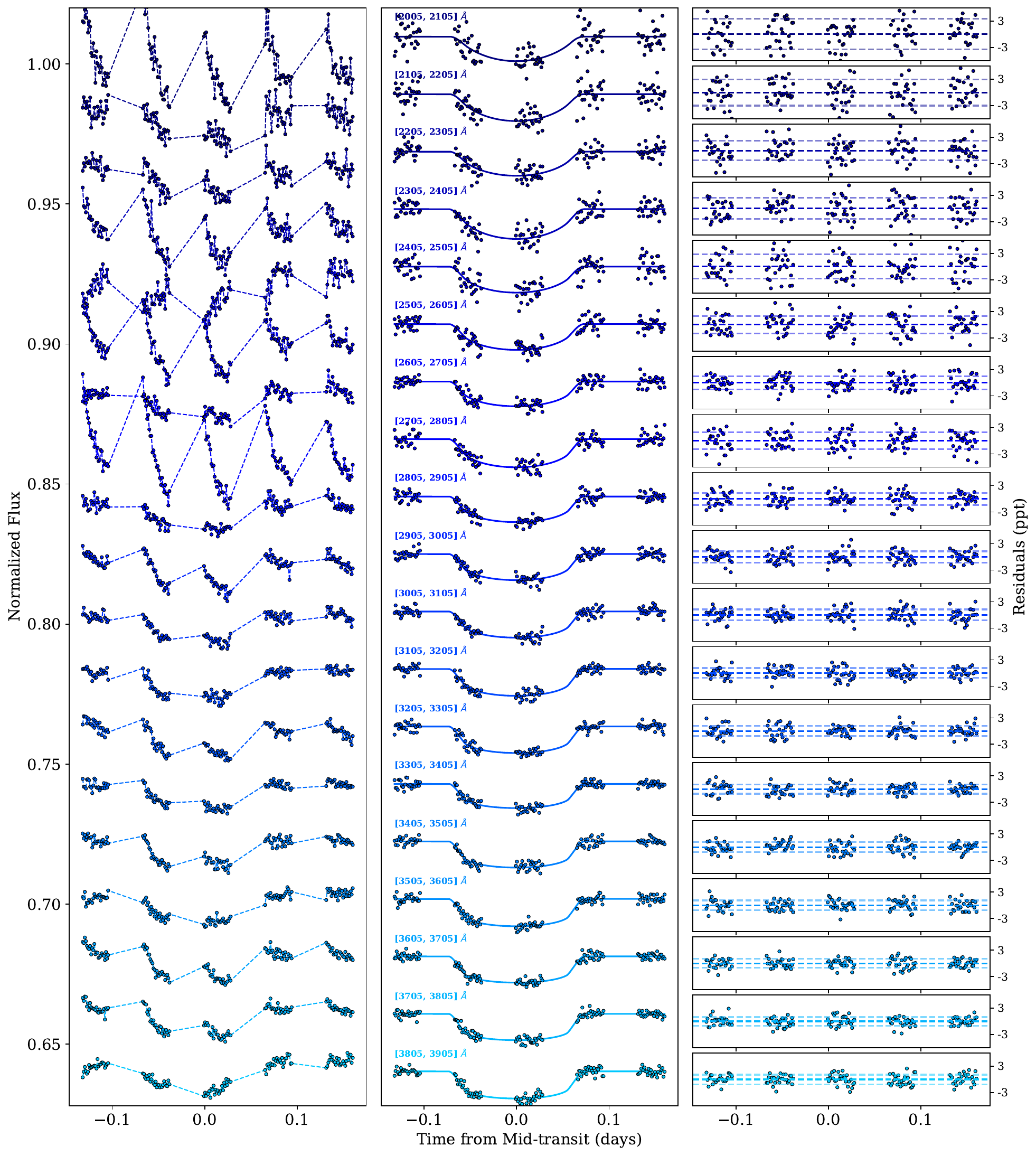}
    \caption{+1 order spectroscopic light curves from 2005 to 3905 \r{A} obtained with the \lluvia pipeline. The left panel shows the raw light curves, the middle panel shows the systematics corrected and model light curves, and the right panel shows the residuals from the corresponding fit.  Light curves are offset vertically by an arbitrary constant for clarity.}
    \label{fig:lcs1}
\end{figure*}

\begin{figure*}
    \centering
    \includegraphics[width = \textwidth]{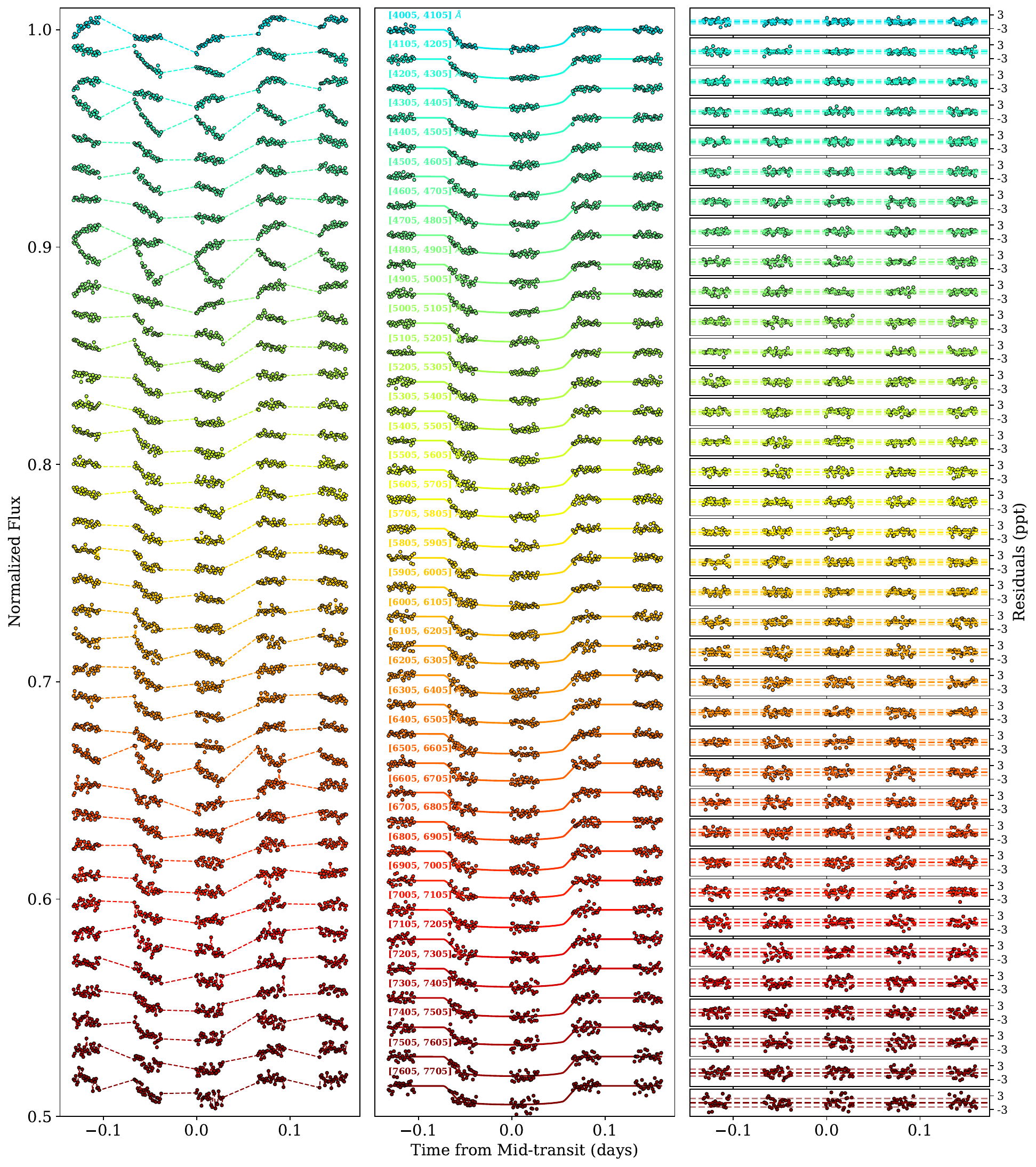}
    \caption{ +1 order spectroscopic light curves from 4005 to 7705 \r{A} obtained with the \lluvia pipeline. The left panel shows the raw light curves, the middle panel shows the systematics corrected and model light curves, and the right panel shows the residuals from the corresponding fit.  Light curves are offset vertically by an arbitrary constant for clarity.}
    \label{fig:lcs2}
\end{figure*}

\begin{figure*}
    \centering
    \includegraphics[width = \textwidth]{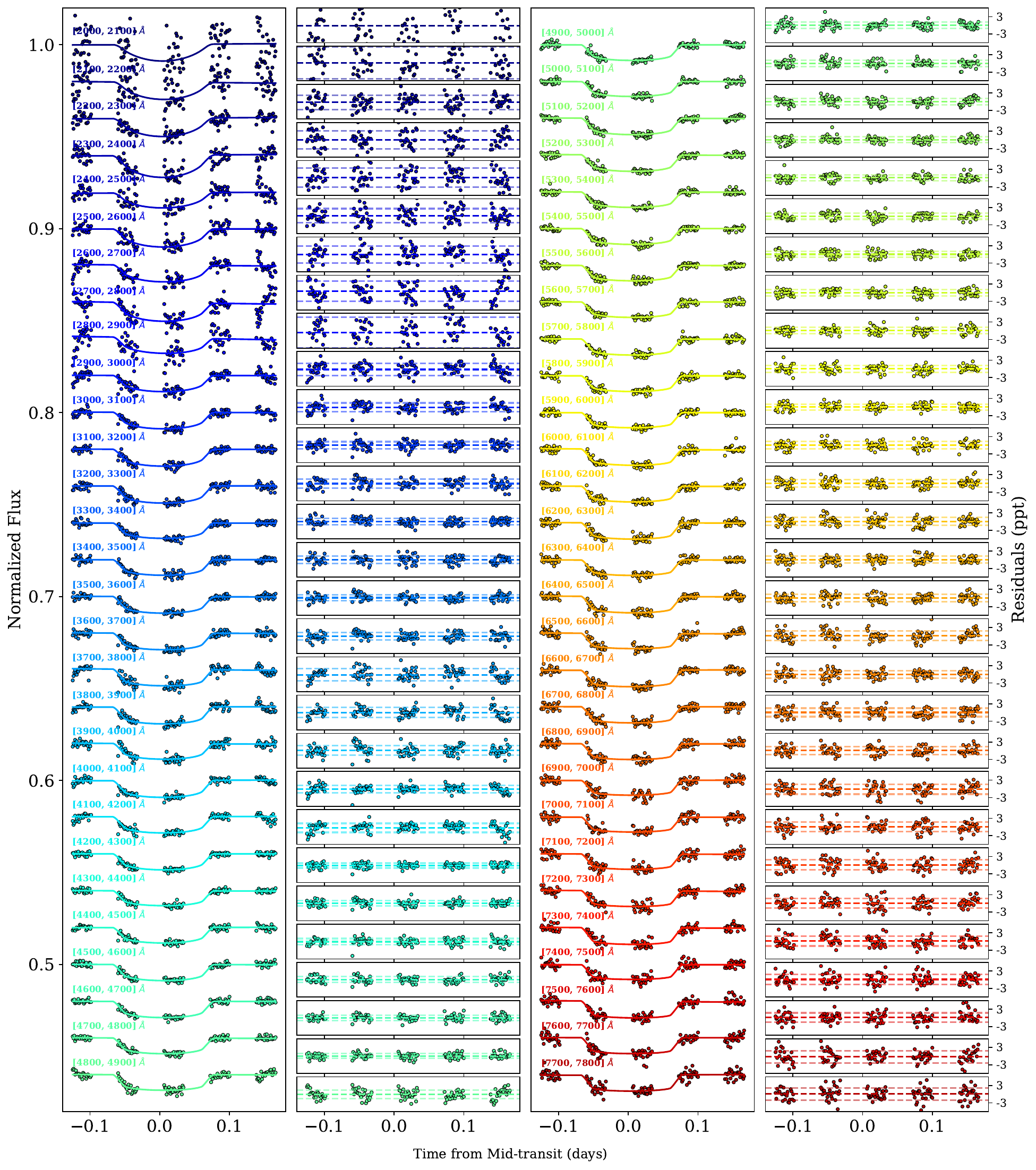}
    \caption{+1 order spectroscopic light curves from 2000 to 7800 \r{A} obtained with the \Hazelnut pipeline.  The first and third columns show the systematics corrected and fitted model light curves, and the second and fourth columns show the corresponding residuals. Light curves are offset vertically by an arbitrary constant for clarity.}
    \label{fig:lcs3}
\end{figure*}

In this section we show the white light curves, and the full set of +1 order spectroscopic light curves extracted with both \lluvia and \Hazelnut pipelines. Figure~\ref{fig:wlc_abby} shows the raw and systematics corrected white light curves, together with the fitted light curve models and residuals, obtained with the \Hazelnut pipeline. Figures \ref{fig:lcs1} and \ref{fig:lcs2} show the spectroscopic light curves obtained with HST WFC3 UVIS G280 using the \lluvia pipeline. Each figure shows the raw light curves (left panel), the systematic corrected and fitted model light curves (middle panel), and the corresponding residuals (right panel). In addition, Figure~\ref{fig:lcs3} shows the spectroscopic light curves calculated with the \Hazelnut pipeline for the +1 and -1 orders, respectively. The first and third columns show the systematics corrected and fitted model light curves, while the second and fourth columns show the corresponding residuals.

\clearpage 
\bibliography{biblio}{}
\bibliographystyle{aasjournal}

\end{document}